\documentclass[longauth]{aa}

\usepackage{graphicx}
\usepackage{txfonts}
\usepackage[colorlinks=true,citecolor=blue,linkcolor=black]{hyperref}
\usepackage{siunitx}
\usepackage{xcolor}
\usepackage{float}
\usepackage{bm}
\usepackage{soul,xcolor}
\setstcolor{red}
\definecolor{alizarin}{rgb}{0.82, 0.1, 0.26}

\newcommand{\hii}{\textup{H}\,\textsc{ii}}
\newcommand{\ha}{\textup{H}\ensuremath{\alpha}}
\newcommand{\hb}{\textup{H}\ensuremath{\beta}}

\newcommand{\SFR}{\ensuremath{\Sigma_{\mathrm{SFR}}}}
\newcommand{\aha}{\ensuremath{A_{\mathrm{ H \alpha}}}}

\begin{document}

   \title{Calibration of hybrid resolved star formation rate recipes based on PHANGS--MUSE \ha\ and \hb\ maps}
    
\author{Francesco~Belfiore\inst{\ref{arcetri}}
        \and
        Adam~K.~Leroy\inst{\ref{ohio}}
        \and
        Jiayi~Sun\inst{\ref{McMaster},\ref{CITA}}
        \and
        Ashley~T.~Barnes\inst{\ref{UBonn}}
        \and
        Médéric~Boquien\inst{\ref{ua}}
        \and
        Yixian~Cao \inst{\ref{mpe}}
         \and
        Enrico~Congiu\inst{\ref{uch}}
        \and
        Daniel~A.~Dale\inst{\ref{wyo}} 
        \and
        Oleg~V.~Egorov\inst{\ref{rechen}}
        \and 
        Cosima~Eibensteiner \inst{\ref{UBonn}}
        \and
        Simon~C.~O.~Glover\inst{\ref{zah}}
        \and
        Kathryn~Grasha \inst{\ref{anu},\ref{a3d}}
        \and   
        Brent~Groves\inst{\ref{uwa}}  
        \and
        Ralf~S.~Klessen\inst{\ref{zah},\ref{zw}}
        \and
        Kathryn~Kreckel\inst{\ref{rechen}}
        \and
        Lukas~Neumann\inst{\ref{UBonn}}
        \and
        Miguel~Querejeta\inst{\ref{oan}}
        \and
        Patricia~Sanchez-Blazquez \inst{\ref{ucm},\ref{iparcos}}
        \and
        Eva~Schinnerer\inst{\ref{mpia}}
        \and
        Thomas~G.~Williams \inst{\ref{mpia}}
    }

\institute{INAF — Osservatorio Astrofisico di Arcetri, Largo E. Fermi 5, I-50125, Florence, Italy\label{arcetri}
   \email{francesco.belfiore@inaf.it}
    \and Department of Astronomy, The Ohio State University, 140 West 18th Avenue, Columbus, OH 43210, USA\label{ohio}
    \and Department of Physics and Astronomy, McMaster University, 1280 Main St. West, Hamilton ON L8S 4M1 Canada\label{McMaster}
    \and Canadian Institute for Theoretical Astrophysics (CITA), University of Toronto, 60 St George Street, Toronto, ON M5S 3H8, Canada\label{CITA}
    \and Argelander-Institut f\"ur Astronomie, Universit\"at Bonn, Auf dem H\"ugel 71, D-53121 Bonn, Germany\label{UBonn}
     \and Centro de Astronomía (CITEVA), Universidad de Antofagasta, Avenida Angamos 601, Antofagasta, Chile\label{ua}
    \and Max-Planck-Institut f\"ur extraterrestrische Physik (MPE), Giessenbachstrasse 1, D-85748 Garching, Germany\label{mpe}
    \and Departamento de Astronom\'{i}a, Universidad de Chile, Camino del Observatorio 1515, Las Condes, Santiago, Chile\label{uch}
    \and Department of Physics and Astronomy, University of Wyoming, Laramie, WY 82071, USA\label{wyo}
    \and Astronomisches Rechen-Institut, Zentrum f\"ur Astronomie der Universit\"at Heidelberg, M\"onchhofstra{\ss}e 12-14, D-69120 Heidelberg, Germany\label{rechen}
    \and Universit\"at Heidelberg, Zentrum f\"ur Astronomie, Institut f\"ur theoretische Astrophysik, Albert-Ueberle-Stra{\ss}e 2, D-69120, Heidelberg, Germany\label{zah}
    \and Research School of Astronomy and Astrophysics, Australian National University, Canberra, ACT 2611, Australia\label{anu}
    \and ARC Centre of Excellence for All Sky Astrophysics in 3 Dimensions (ASTRO 3D), Australia\label{a3d}
    \and  International Centre for Radio Astronomy Research, University of Western Australia, 7 Fairway, Crawley, 6009, WA, Australia\label{uwa}
    \and Universit\"at Heidelberg, Interdisziplin\"ares Zentrum f\"ur Wissenschaftliches Rechnen, Im Neuenheimer Feld 205, D-69120 Heidelberg, Germany\label{zw}
    \and Observatorio Astronómico Nacional (IGN), C/Alfonso XII, 3, E-28014 Madrid, Spain\label{oan}   
    \and Departamento de F\'isica de la Tierra y Astrof\'isica, Universidad Complutense de Madrid, E-28040 Madrid, Spain \label{ucm}
    \and Instituto de F\'{\i}sica de Part\'{\i}culas y del Cosmos IPARCOS, Facultad de CC Físicas, UCM, E-28040, Madrid, Spain \label{iparcos}
    \and Max-Planck-Institute for Astronomy, K\"onigstuhl 17, D-69117 Heidelberg, Germany\label{mpia}
}

\date{Received XXX; accepted XXX}

 
\abstract{
Mapping star-formation rates (SFR) within galaxies is key to unveiling their assembly and evolution. Calibrations exist for computing SFR from a combination of ultraviolet and 
infrared bands for galaxies as integrated systems, but their applicability to sub-galactic (kpc) scales remains largely untested. Here we use integral field spectroscopy of 19 nearby ($D <$ 20 Mpc) galaxies obtained by PHANGS--MUSE to derive accurate Balmer decrements (H$\alpha$/H$\beta$) and attenuation-corrected H$\alpha$ maps. We combine this information with mid-infrared maps from \textit{WISE} at 22 $\rm \mu m$, and ultraviolet maps from \textit{GALEX} in the far-UV band, to derive SFR surface densities in nearby galaxies on resolved (kpc) scales. Using the \ha\ attenuation-corrected SFR as a reference, we find that hybrid recipes from the literature overestimate the SFR in regions of low SFR surface density, low specific star-formation rate (sSFR), low attenuation and old stellar ages. We attribute these trends to heating of the dust by old stellar populations (`IR cirrus'). We calibrate this effect by proposing functional forms for the coefficients in front of the IR term which depend on band ratios sensitive to the sSFR. Such recipes return SFR estimates in agreement with those in the literature at high sSFR ($\rm log(sSFR/yr^{-1}) > -9.9$). Moreover, they lead to negligible bias and $<$0.16 dex scatter when compared to our reference attenuation-corrected SFR from \ha. These calibrations prove reliable as a function of physical scale. In particular, they agree within 10\% with the attenuation corrections computed from the Balmer decrement on 100~pc scales. Despite small quantitative differences, our calibrations are also applicable to integrated galaxy scales probed by the MaNGA survey, albeit with a larger scatter (up to 0.22~dex). Observations with \textit{JWST} open up the possibility to calibrate these relations in nearby galaxies with cloud-scale ($\sim$100 pc) resolution mid-IR imaging.}
\titlerunning{SFR Calibrations Based on Resolved Balmer Decrements}
\authorrunning{F. Belfiore}
\maketitle


\section{Introduction}
\label{sec:intro}

Estimates of the recent star-formation rate (SFR) play a central role in studies of galaxy evolution, chemical enrichment, stellar feedback, and the physics of stellar birth. Estimating the SFR in distant galaxies entails measuring light from young massive stars, either directly (e.g.\ in the ultraviolet, UV), or indirectly, via recombination line emission (e.g. \ha) or the radiation re-emitted by dust in the infrared (IR).  

From the cosmological perspective, for example, knowledge on how the dust attenuation affects UV emission is fundamental to correctly reconstruct the  peak and decline of the cosmic star formation history at $z < 4$ \citep{Lilly1996, Gruppioni2013, Madau2014, Koprowski2017}, and interpret the properties of the first galaxies, which may already be subject to substantial dust obscuration \citep{Casey2018, Gruppioni2020, Algera2022}. The physics of the star formation process, on the other hand, can be probed in the extragalactic context by studying the so-called `star formation laws', relating gas content with SFR for entire galaxies or sub-galactic regions \citep{Kennicutt1998, Bigiel2008, Bacchini2019}. Accurate dust corrections have been instrumental in reducing the systematic uncertainties in the measurement of slope of these relations \citep{Kennicutt2021}, and therefore in discriminating between competing physical models \citep{Sun2022}.

Unfortunately, both UV and IR emission can arise from sources other than young stars, and different SFR tracers are dependent on the age of the stellar population and other physical properties of the interstellar medium (ISM) in complex ways. As a result, calibrating SFR estimates and linking between different systems remains an important, open topic \citep[for reviews see][]{Kennicutt1998,Murphy2011, Wuyts2011, Kennicutt2012,Calzetti2013, Boquien2016}.

In this paper, we focus on measurements of the SFR per unit area (\SFR) on kiloparsec scales, and attempt to link three of the most commonly used calibration methods: (1) \ha\ emission corrected for dust attenuation by using the Balmer decrement \citep[e.g.][and posited as early as \citealt{Berman1936}]{Groves2012a}; (2) \ha\ emission corrected using an empirically-calibrated combination with \textit{WISE} 22 $\rm \mu m$ (or \textit{Spitzer} 24$\mu$m, \textit{IRAS} 25$\mu$m) emission \citep[e.g.][]{Kennicutt2007,Calzetti2007}; and (3) UV emission combined with \textit{WISE} 22 $\rm \mu m$ emission \citep[e.g.][]{Meurer1999,Leroy2008,Hao2011, Boquien2016}.

The first technique contrasts the expected ratio of \ha\ to \hb\ line emission for Case B recombination with the observed ratio to calculate the ratio of attenuations. Combined with an adopted attenuation curve, the paired measurement of \ha\ and \hb\ line emission yields the attenuation affecting \ha, \aha, and the attenuation-corrected \ha\ flux, $F_{\rm H\alpha}^{\rm corr}$. Although this method has a long pedigree, it has been used more widely in recent years thanks to the broad availability of optical integral field spectroscopy (IFS) from surveys such as CALIFA \citep{Sanchez2012}, SAMI \citep{Croom2012}, and MaNGA \citep{Bundy2015}.

The second method consists of combining \ha\ with mid-IR maps, generally (but not always) via a linear combination. The ratio of mid-IR to \ha\ emission is treated as an indicator of \aha\ or, put another way, the mid-IR emission is used to trace the portion of \ha\ emission absorbed by dust. Although this approach can be theoretically motivated, its calibration is fundamentally empirical. Namely, the mid-IR flux is scaled by a (pre)factor set by benchmarking against some `gold standard' reference SFR tracer, most commonly attenuation-corrected hydrogen recombination lines as described in the previous paragraph \citep[e.g.][]{Calzetti2007, Kennicutt2009}.

The third approach combines UV and IR emission. We consider linear combinations of individual bands \citep[e.g.][]{Thilker2007,Leroy2008,Hao2011, Leroy2012, Boquien2016, Leroy2019}, but previous versions of this approach also include multi-band or higher-order calibrations, and summing full bolometric estimates of the IR and UV emission \citep[e.g.][]{Meurer1999}. In a broad sense, this approach represents a simplified adaptation of the increasingly common practice of fitting population synthesis models to the spectral energy distribution (SED) of galaxies in the UV, optical, and IR bands \citep[e.g.][]{daCunha2008, Conroy2013a,Salim2016,Salim2018, Nersesian2019, Hunt2019}.

We focus here on the linear combination of UV and the $22~\mu$m mid-IR band. The mid-IR offers several advantages with respect to the far-IR. 
Physically, the mid-IR continuum is expected to be directly sensitive to the incident radiation field.
Empirically, the warm mid-IR dust appears to be physically associated with regions of massive star formation \citep[e.g.][]{Helou2004,Relano2010} with a smaller contribution from cool dust heated by older stellar populations than the far-IR. Crucially, \textit{Spitzer} had excellent performance at $24~\mu$m, and more recently \textit{WISE} \citep{Wright2010} provided mid-IR data at sufficient resolution ($\theta_{\rm FWHM} \approx 12\arcsec$ at $22~\mu$m) to map nearby star-forming galaxies \citep{Jarrett2013}. Further emphasizing the importance of the mid-IR, the \textit{James Webb Space Telescope} (\textit{JWST}) is providing maps of the mid-IR emission at even higher resolution and sensitivity for both local and distant galaxies.


Despite the utility of the mid-IR based methods, their calibration remains exclusively empirical. Fortunately, studies of nearby galaxies have established self-consistent calibrations for the mid-IR in combination with UV or H$\alpha$ to estimate the SFR for whole galaxies \citep{Kennicutt2009, Hao2011, Catalan-Torrecilla2015, Leroy2019}. Another line of studies have established calibrations that can be applied to individual star-forming regions inside of galaxies \citep[including][]{Kennicutt2007,Calzetti2007,Murphy2011,Boquien2016}. These calibrations are generally found to be accurate to $\sim$ 0.1--0.15 dex \citep{Kennicutt2009, Leroy2012}, but systematically differ from each other at the level of $\sim$ 0.2 dex.

Physically, mid-IR emission reflects the combined effect of dust abundance, relative geometry of dust and stars, and dust heating by the surrounding stellar populations. While young stars may provide the bulk of the heating in individual star-forming regions, in more quiescent regions older stars also play a substantial role in heating the dust and powering the emission in the IR \citep[e.g.][]{Groves2012,Li2013}. High-resolution studies ($\lesssim$ 100~pc) can isolate individual regions and effectively avoid or filter out this diffuse background or `infrared cirrus' \citep[e.g.][]{Calzetti2007}. However, on the larger scales of kiloparsecs or whole galaxies, this IR cirrus will be an unavoidable part of the measurement \citep{Boquien2016}. This fact leads to a non-linear relation between total infrared luminosity  and the SFR that is dependent on the specific SFR ($\rm sSFR = SFR/M_\star$, e.g. \citealt{Cortese2008, Boquien2021}). The contribution of the IR cirrus may be physically modeled and subtracted \citep[e.g.][]{Leroy2012}. Or the calibration of the IR term may take into account the contribution of this diffuse dust component \citep{Boquien2016}.

So far, the lack of quality reference maps has been a major obstacle to calibrating the IR term in SFR prescriptions to apply at sub-galactic scales. In this work, we revisit these questions prompted by a unique combination of new data. The PHANGS--MUSE survey \citep{Emsellem2022} has recently obtained comprehensive integral field spectroscopy (IFS) mapping of the star-forming discs of $19$ nearby galaxies (distance $<20$~Mpc), using the MUSE (Multi-unit Spectroscopic Explorer) instrument \citep{Bacon2010} on the ESO Very Large Telescope. These observations give us access to high signal-to-noise maps of \ha\ and \hb, free from the continuum subtraction systematics associated with narrow-band images. Because PHANGS--MUSE targets nearby galaxies, the relatively coarse $\sim 5{-}15\arcsec$ resolution of mid-IR maps from \textit{WISE} or \textit{Spitzer} translates to $\lesssim 1$~kpc. This scale is fortuitously set, since it allows us to define regions with significant numbers of the relatively rare massive stars, yet not so large as to merge multiple adjacent star-forming regions. 

Extensive, homogenised multi-wavelength imaging is available for the PHANGS--MUSE targets from \textit{GALEX} far-UV (FUV, $\lambda \sim 154$~nm) and near-UV (NUV, $\lambda \sim 231$~nm), as well as \textit{WISE} Band 1, 2, 3, and 4 imaging ($\lambda = 3.4$, $4.5$, $12$, and $22~\mu$m) with 15$\arcsec$ or $\sim 1$~kpc resolution \citep{Leroy2019}. Such a data set allows us to check for consistency among the three methods of estimating SFR discussed above, quantify systematic uncertainties, and derive new empirical coefficients appropriate for this scale and sample, using the attenuation-corrected \ha\ maps from MUSE as a reference. 

A key motivation for this new calibration effort is the desire to derive SFR maps that are anchored in the Balmer-corrected MUSE data and combine \textit{WISE} IR maps with new narrow-band \ha\ observations of nearby galaxies (Razza et al., in preparation). Moreover, we wish to integrate the UV+IR-based SFR maps widely available for the nearest galaxies ($D \lesssim 50$~Mpc, e.g. \citealt{Jarrett2013, Leroy2019}) into a common framework with the attenuation-corrected \ha\ derived from optical IFS mapping that has been obtained for large samples of more distant ($D \sim 20{-}150$~Mpc) galaxies \citep{Belfiore2018, Medling2018, Sanchez2020}. Finally, observations in the mid-IR with \textit{JWST} now provide us with a high-resolution view of embedded star formation in the nearby Universe, therefore making it urgent to verify the validity of SFR calibrations across a range of physical scales.



With these goals in mind, Section \ref{sec:data} presents our compilation of $15\arcsec$ resolution MUSE, UV, and IR measurements, and describes our methods and assumptions. Section \ref{sec:results} compares results for \SFR\ and \aha\ calculated using previous calibrations and presents updated recipes for kpc-scale SFR calibrations using \ha+IR and UV+IR. In Sect. \ref{sec:discussion}, we discuss the impact on our new calibrations together with their limitations. We summarise our results in Sect. \ref{sec:summary}.


\section{Data and methods}
\label{sec:data}

The starting point of this work is the sample of 19 galaxies targeted by the PHANGS-MUSE survey. These objects were selected to be nearby ($D < 20$~Mpc), close to the star formation main sequence, and moderately inclined ($i < 57.3^\circ$). The main properties of the sample, which spans the stellar mass range $\log(M_\star/M_{\odot}) = 9.4 {-} 11.0$, are presented in \cite{Emsellem2022} and summarised in Table \ref{tab:sample}.

\begin{table}
\caption{Key properties of the PHANGS--MUSE sample of galaxies used in this work. Distances are taken from the compilation of \cite{Anand2021}. Stellar masses, SFR, and offset from the star-formation main sequence ($\Delta_{\rm SFMS}$) are taken from the PHANGS sample paper \citep{Leroy2021a}, inclination ($i$) are taken from \cite{Lang2020}. $\rm \theta_{15}$ is the size in kpc of the 15$''$ resolution element used in this analysis.}
\label{tab:tab1}
\centering
\begin{tabular}{lccccc}
\hline \hline
Name & Distance & ${\rm Log}(M_\star) $ & $\rm Log(SFR) $  & $i$ & $\rm \theta_{15}$ \\
     & Mpc & [$M_\odot$]	& [$M_\odot~\mathrm{yr}^{-1}$]	& deg  & kpc \\
\hline
\hline
NGC~0628 & 9.8 & 10.34  & 0.18 &  8.9 & 0.7 \\
NGC~1087 & 15.9 & 9.93  & 0.33 & 42.9 & 1.2 \\
NGC~1300 & 19.0 & 10.62  & $-$0.18 & 31.8 & 1.4 \\
NGC~1365 & 19.6 & 10.99  & 0.72 & 55.4 & 1.4\\
NGC~1385 & 17.2 & 9.98  & 0.50 & 44.0 &  1.3 \\
NGC~1433 & 18.6 & 10.87  & $-$0.36 &  28.6 & 1.4 \\
NGC~1512 & 18.8 & 10.71  & $-$0.21 & 42.5 & 1.4 \\
NGC~1566 & 17.7 & 10.78  & 0.29  & 29.5 & 1.3 \\
NGC~1672 & 19.4 & 10.73  & 0.56 & 42.6 & 1.4 \\
NGC~2835 & 12.2 & 10.00  & 0.26 & 41.3 & 0.9 \\
NGC~3351 & 10.0 & 10.36  & 0.05  & 45.1 & 0.7 \\
NGC~3627 & 11.3 & 10.83 & 0.19 & 57.3 & 0.8 \\
NGC~4254 & 13.1 & 10.42  & 0.37 & 34.4 & 1.0 \\
NGC~4303 & 17.0 & 10.52  & 0.54 & 23.5 & 1.2\\
NGC~4321 & 15.2 & 10.75  & 0.21  & 38.5 & 1.1\\
NGC~4535 & 15.8 & 10.53 & 0.14 &  44.7 & 1.1\\
NGC~5068 & 5.2 & 9.40  & 0.02 & 35.7 & 0.4 \\
NGC~7496 & 18.7 & 10.00  & 0.53 &  35.9 & 1.4\\
IC~5332 & 9.0 & 9.67  & 0.01 & 26.9 & 0.7\\
\hline
\end{tabular}
\label{tab:sample}
\end{table}

We compiled and analysed a data set that includes intensities and associated uncertainties at a $15\arcsec$ common resolution from MUSE, \textit{GALEX}, and \textit{WISE}. The limiting resolution is set by WISE W4 band centered at $22~\mu$m. For the distances of our galaxy sample ($D=5.2-19.6$~Mpc), this angular resolution corresponds to a median physical resolution of 1.1~kpc, varying across our sample from 400~pc to 1.4~kpc (see Table \ref{tab:sample}).


\subsection{PHANGS--MUSE, optical IFS}
\label{sec:MUSE}

Each galaxy in the PHANGS--MUSE sample was observed with several (three to fifteen) MUSE pointings, each $1\arcmin \times 1\arcmin$. Observing strategy, data reduction and analysis of this data set are described in detail in \cite{Emsellem2022}. The data reduction was carried out via \textsc{pymusepipe}, a python wrapper to the ESOREX MUSE reduction recipes \citep{Weilbacher2020a}. 
The point spread function (PSF) and sky background of each MUSE pointing were determined by comparison with wide-field $R$-band imaging (Razza et al., in preparation). Reconstructed photometry from MUSE is consistent with Sloan Digital Sky Survey (SDSS) imaging with a typical scatter of 0.04~mag. In this work we mostly use datacubes where the native-resolution mosaicked cube was convolved with a suitable kernel in order to deliver a common Gaussian resolution of 15$\arcsec$ full width at half maximum (FWHM) and spaxels 2\farcs4 in size. In Sect. \ref{sec:ha_narrow} we study the effect of spatial resolution on our results and we therefore use the `optimally convolved' (copt) datacubes, described in \cite{Emsellem2022}. In these data products the PSF was homogenised across each individual mosaic and as a function of wavelength using the \textsc{pypher} tool \citep{Boucaud2016} to obtain, for each galaxy, a best possible common Gaussian PSF. The median PSF FWHM of the copt data is 0\farcs95.

Emission-line fluxes and other stellar population properties were obtained by using the PHANGS data analysis pipeline (\textsc{dap}), as described in \citet{Emsellem2022}. The \textsc{dap} was used to perform three full spectral fitting steps: the first optimised for the recovery of the stellar kinematics, the second aimed at determining the stellar population properties, and the final one optimised for extraction of fluxes and kinematics of gas emission lines. The core of all three spectral fitting modules is the \textsc{pPXF} python module \citep{Cappellari2004, Cappellari2017}. The stellar continuum was fitted using a set of simple stellar population models from the E-MILES library \citep{Vazdekis2012}. 
From the \textsc{dap} output we use maps of \ha\ and \hb\ line emission (corrected for Galactic extinction), and also consider the mean light-weighted age of the stellar population in Sect. \ref{sec:secondary_dep}. We applied a set of custom-designed masks to the output MUSE maps. In particular, for the 15$\arcsec$-resolution data we masked foreground stars brighter than 16.5~mag in Gaia G-band from Gaia Data Release 2 \citep{GaiaCollaboration2018} and applied a few additional masks to stars which were missed and the central regions of galaxies hosting active galactic nuclei (AGN).

\subsection{Mid-IR and UV maps}
\label{sec:z0mgs}

IR and UV imaging was obtained from z0MGS ($z=0$ multi-wavelength galaxy synthesis, \citealt{Leroy2019}), an atlas of \textit{WISE} and \textit{GALEX} \citep{Martin2005} images of nearby ($D<$ 50 Mpc) galaxies. z0MGS is based on the unWISE reprocessing by \citet{Lang2014} and draws heavily on HyperLEDA \citep{Makarov2014} for the atlas construction. 

Images used in this work were taken from the publicly available atlas, and were convolved to a common 15$\arcsec$ FWHM resolution, with a pixel size of 5\farcs5. Masking based on GAIA DR2, 2MASS \citep{Skrutskie2006}, and HyperLEDA was carried out to remove foreground stars or other galaxies in the field, and a local background subtraction was performed for each object. \textit{GALEX} imaging was corrected for Galactic extinction following \cite{Peek2013} and the \cite{Schlegel1998} dust maps. 
Stellar mass surface density maps were derived according to the prescription of \cite{Leroy2019}, who used the \textit{WISE} $\rm 3.4 \mu m $ W1 band in combination with a mass-to-light ratio that scales with SFR to obtain masses in agreement with \cite{Salim2018}. As already noted in \cite{Leroy2021a}, this mass estimate agrees well with the masses derived from full spectral fitting of the MUSE data, up to a roughly constant multiplicative offset, which makes the MUSE masses larger by 0.09~dex. 

Since much of the work on SFR calibration in the literature is based on the \textit{Spitzer} MIPS 24 $\rm \mu m$ band, we have compared the \textit{WISE} W4 maps from z0MGS with \textit{Spitzer} MIPS 24 $\rm \mu m$ mosaics for a subset of 8 galaxies with available \textit{Spitzer} images from either the SINGS \citep{Kennicutt2003} or LVL \citep{Dale2009} programs (NGC~0628, NGC~1512, NGC~1566, NGC~3351, NGC~3627, NGC~4254, NGC~4321, NGC~5068). The \textit{Spitzer} MIPS 24 $\rm \mu m$ images were convolved to the same common 15$\arcsec$ FWHM resolution. While the \textit{Spitzer} maps are substantially deeper than the \textit{WISE} ones (by a factor of 6 on average), the fluxes in regions which are detected in both maps ($S/N >$ 5 in the \textit{WISE} maps) are in excellent agreement. In particular, we find a small multiplicative offset, which goes in the direction of making the MIPS flux densities smaller by 11\% (0.05~dex), in line with the previous comparisons \citep{Jarrett2011,Jarrett2013, Brown2014}.\footnote{The \textit{WISE} data used here are calibrated to $\rm MJy\,sr^{-1}$ units following the WISE documentation Sect. 4.4h, and \cite{Wright2010} assuming a $ \propto \nu^{-2}$ reference spectrum with no colour-corrections. \cite{Brown2014} show that the W4 colour correction, which depends on the spectral index of the source, can be significant. However, we do not perform such a correction because because we wish to directly compare with \cite{Leroy2019} and \cite{Salim2018}, who do not perform such a correction to their W4 fluxes, and because we we wish to derive prescriptions that can be applied to the WISE data without such preprocessing. In any case, the SED fitting necessary to determine it is outside the scope of this work.} In energy units, which are used in the next section to define the SFR calibration, $\log{(\nu_{\mathrm{MIPS}} F_{\nu, \mathrm{MIPS}})} =\log{(\nu_{\mathrm{W4}} F_{\nu, \mathrm{W4}})} -0.08$, with an uncertainty due to galaxy-to-galaxy scatter of 0.03 dex.

\begin{figure*}
	\centering
	\includegraphics[width=0.9\textwidth, trim=20 60 20 20, clip]{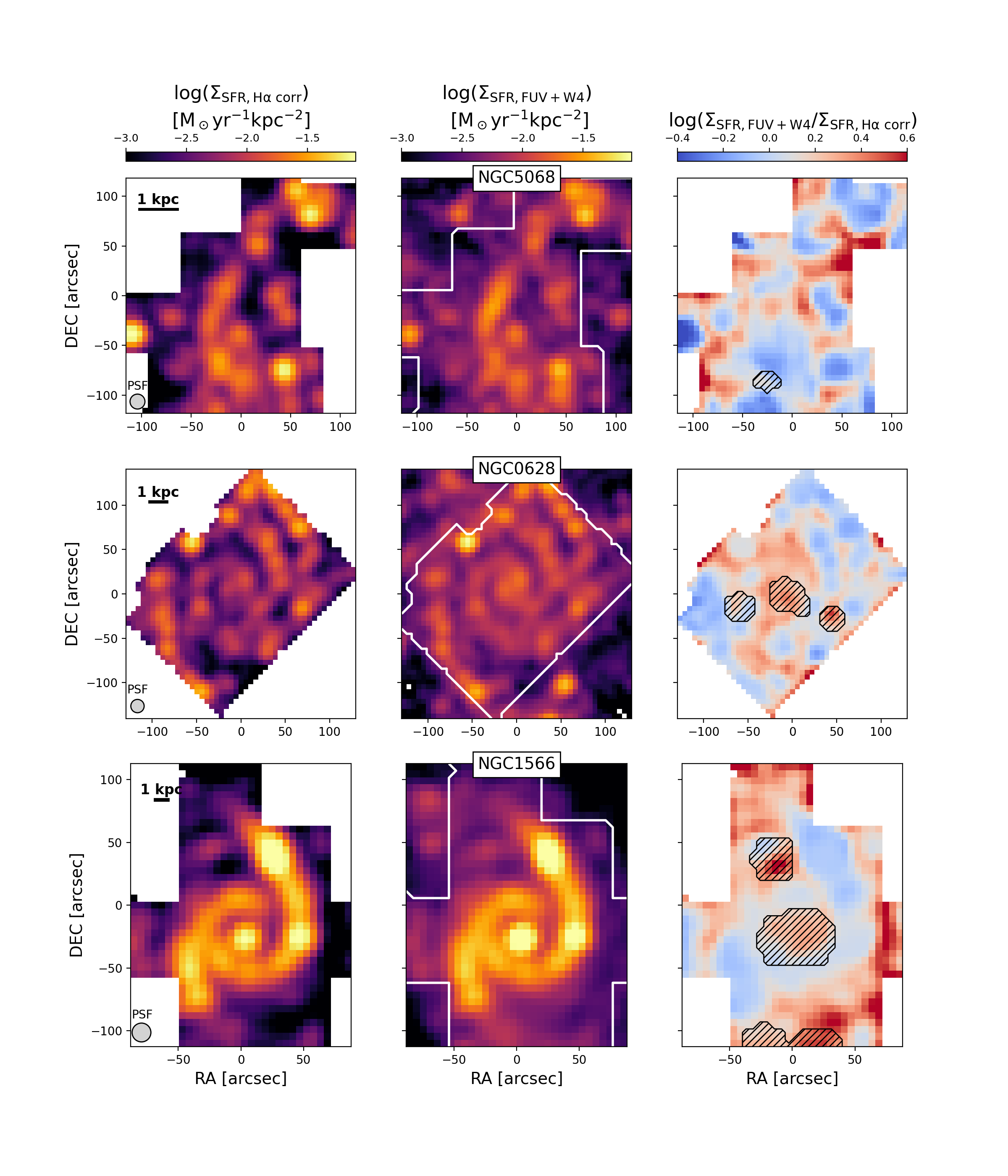}
	\caption{Comparison between $\rm \Sigma_{SFR}/(M_\odot \ kpc^{-2} \ yr^{-1})$ estimates at 15$\arcsec$ resolution from extinction-corrected \ha\ emission and UV+IR (FUV + W4) using the \cite{Leroy2019} prescription for three example galaxies. The maps are shown on the original pixel size of 5\farcs5, even though the analysis in this paper is based on sampling the data at $\sim$ 15$\arcsec$, and are scaled to the same level for ease of comparison. The white contours in the UV+IR-based map represent the boundary of the MUSE mosaics. The right column shows the difference between the two SFR estimates in dex (the average uncertainty in the measured ratio due to the error in the fluxes is 0.03~dex). The hatched regions represent masked areas, namely regions contaminated by foreground stars and the AGN in NGC~1566.}
	\label{fig:fig1}
\end{figure*}

\subsection{Data processing}
In order to generate a catalog of distinct regions at our working resolution, the \ha\ and \hb\ 15$\arcsec$-resolution maps are resampled with pixels of 14\farcs4 (comparable to the 15$\arcsec$  FWHM). The z0MGS data are reprojected (using the \textsc{reproject} python module) onto the same world coordinate system and their pixel size is matched to that of the MUSE data. A signal-to-noise cut of 5 is applied to the \ha\ and \hb\ lines, but this only affects a negligible fraction of our data set ($<3$\% of the regions). This procedure leads to a sample of 1759 regions, covering a total area of $\rm \sim 100 \ arcmin^2 $ or $\rm 10^4 ~ kpc^2$ across our $19$ targets. 

Stellar and SFR surface density measurements are corrected for the effect of inclination by multiplication by cosine of the inclination angles compiled in \citealt{Leroy2021a}.

\begin{figure*}
	\centering
	\includegraphics[width=1\textwidth, trim=20 40 20 0, clip]{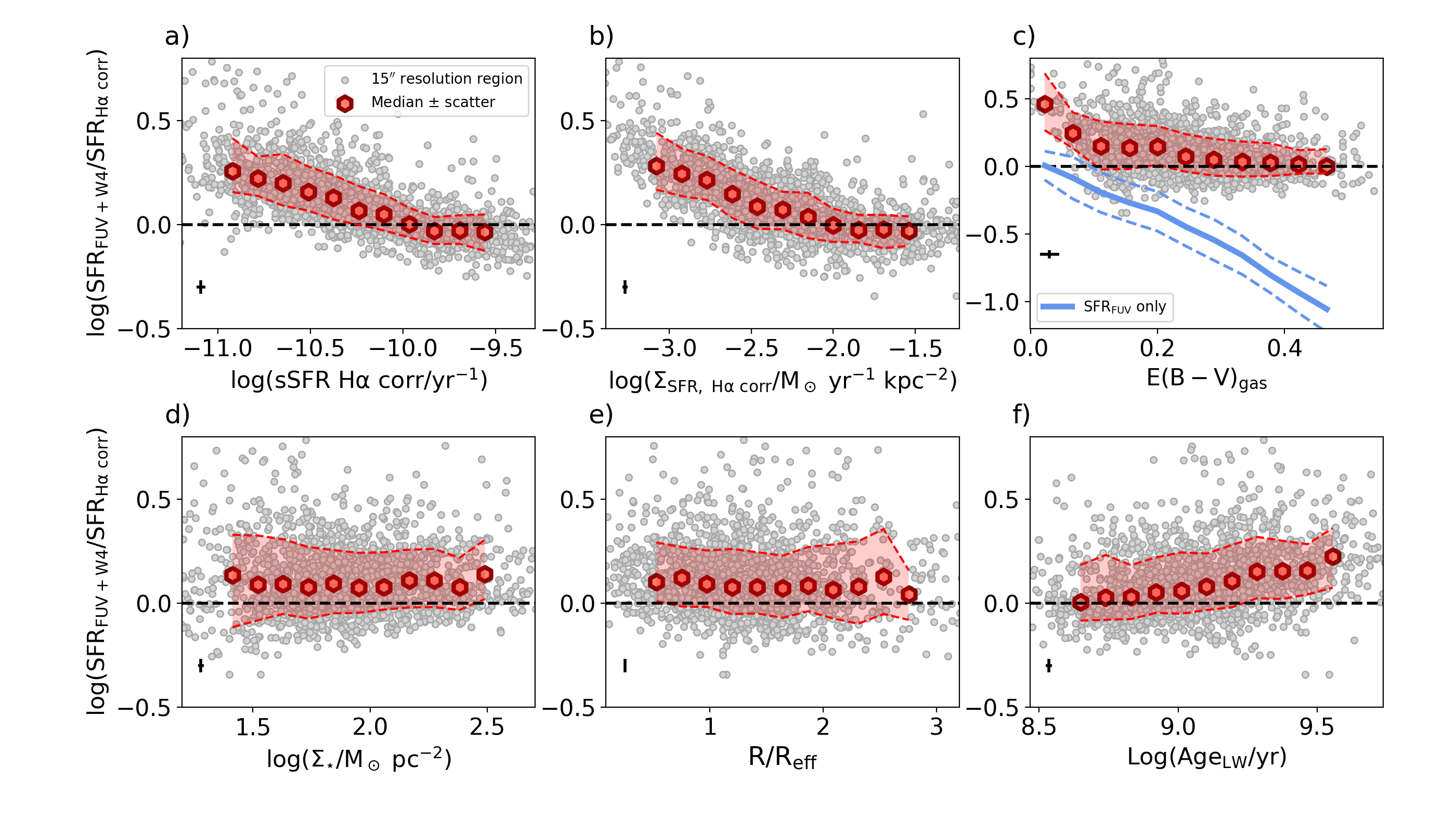}
	\caption{Dependence of the ratio between SFR measured from FUV+W4 (and the \citealt{Leroy2019} coefficient) and that measured from attenuation-corrected \ha\ on various physical properties for a sample of 1759 $\sim$ kpc-sized regions taken from the 19 galaxies of the PHANGS-MUSE sample: a) sSFR, b) SFR surface density (both measured from attenuation-corrected \ha), c) $E(B-V)$ of the ionised gas measured from the Balmer decrement, d) stellar mass surface density, e) deprojected galactocentric radius normalised to $\rm R_{eff}$, f) light-weighted mean age of the stellar population, derived from fits to the MUSE stellar continuum,. The grey points represent individual $\sim$ 15$\arcsec$ regions, while the red hexagons and the dashed red lines represent the median and scatter ($\rm 16^{th}-84^{th}$ percentiles) of the distribution. In Panel c) we also show the result of calculating the SFR from the FUV flux, without considering the IR term (blue lines). This demonstrates that the inclusion of the IR term correctly reproduces the \ha-based SFR for highly attenuated regions, but it overestimates the attenuation level for those regions with low Balmer decrement. These trends are consistent with an increasing contamination from old stellar population to the dust heating when moving to lower SFR levels.}
	\label{fig:fig2}
\end{figure*}

\begin{figure*}
	\centering
	\includegraphics[width=1\textwidth, trim=20 40 20 15, clip]{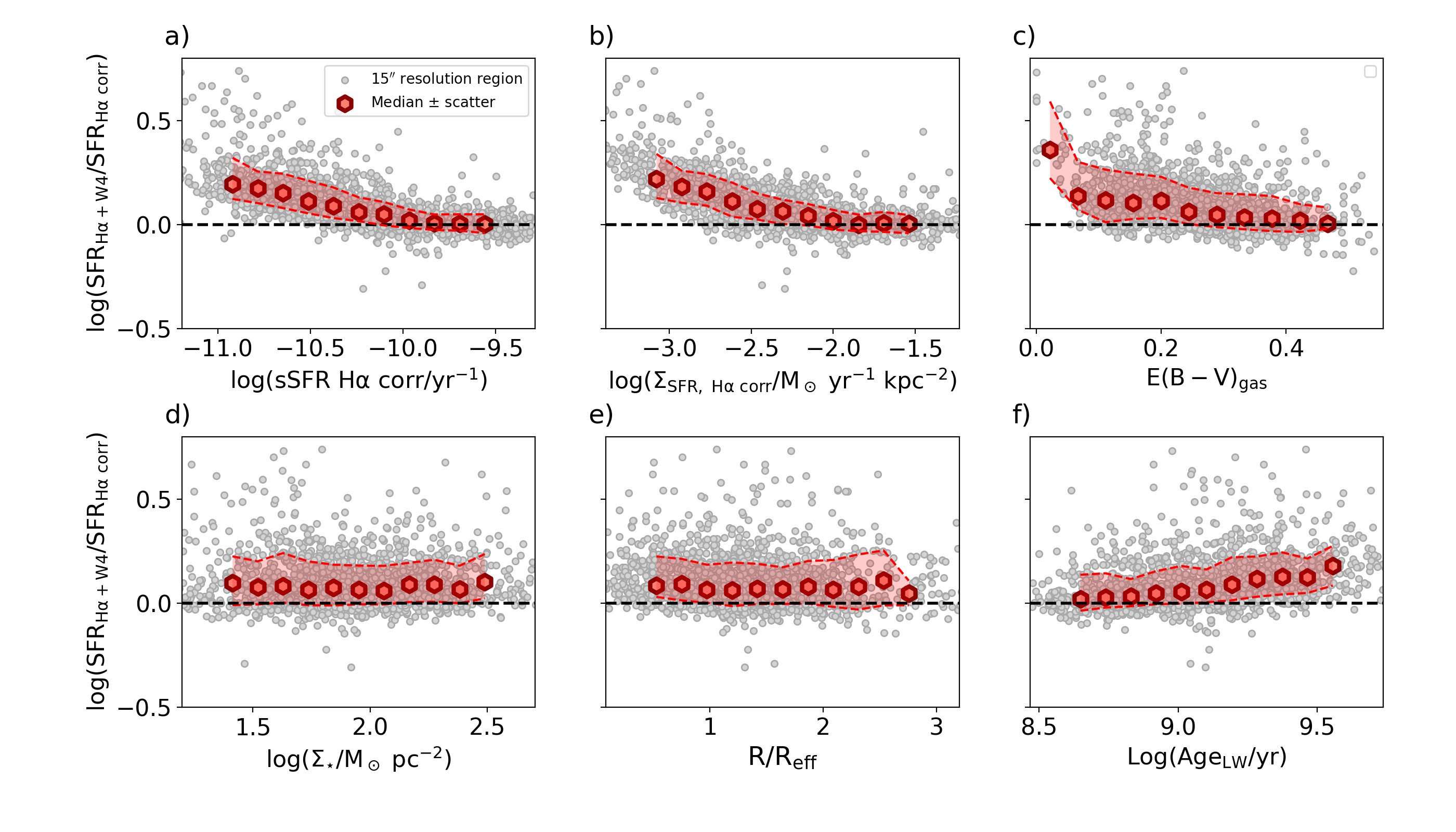}
	\caption{Same as Fig. \ref{fig:fig2} but using \ha+W4 instead of FUV+W4. The coefficient $C_{\rm W4}^{\rm H\alpha}$ is taken from \cite{Calzetti2007}.}
	\label{fig:appA1}
\end{figure*}

\subsection{Formalism for computing SFR}
\label{sec:formalism}

The attenuation-corrected \ha\ flux is obtained as
\begin{equation}
L_{\rm H\alpha, corr} = L_{\rm H\alpha} 10 ^{0.4 ~  A_{\rm H\alpha} }  = L_{\rm H\alpha} 10 ^{0.4~ k_{\rm H\alpha}~ E(B-V)},
\label{eq:ext_corr_ha}
\end{equation}
where $k_{\rm H\alpha}$ is the value of the reddening curve at the \ha\ wavelength. We quote all luminosities in energy units ($\rm erg \, s^{-1}$), i.e. $L \equiv \nu L_\nu$, where $\nu$ is the (effective) frequency of the band considered. 

We measure $E(B-V)$ from the Balmer decrement (BD) assuming Case B recombination, temperature $T=10^4$ K and density $n_{\rm e} = 10^2 \, \mathrm{cm}^{-2}$, leading to  $L_{\rm H\alpha, corr}/L_{\rm H\beta, corr} = 2.86$. This ratio is almost independent of density for the range of densities relevant to \hii\ regions, and has only a small temperature dependence \citep{Osterbrock2006}. Under these assumptions 
\begin{equation}
E(B-V)_{\rm BD} = \frac{2.5}{k_{\rm H\beta} - k_{\rm H\alpha}} \log_{10} \left[ \frac{  L_{\rm H\alpha}/L_{\rm H\beta}  } {2.86} \right].
\label{ebv}
\end{equation}
On the scales probed by our observations dust will be mixed with the gas, leading to the need to consider an effective attenuation law, and not just the effect of dust extinction along the line of sight. 
We nonetheless use a foreground screen attenuation model with total-to-selective extinction $R_V = 3.1$ equivalent to the Milky Way extinction law of \cite{O'Donnell1994}, which represents a small modification to the one proposed by \cite{Cardelli1989}. There is some observational support for the use of such a simplified foreground screen model as applied to the nebular lines in galaxies \citep{Calzetti1996, Wild2011a}. Moreover, the use of a Milky-Way-like extinction law follows the conventions of the field, making our results directly comparable with the literature \citep{Calzetti2000, Kennicutt2009, Boquien2019}. We discuss the effect of making different assumptions regarding the attenuation curve for the nebular component in Sect. \ref{sec:discussion}.

The conversion factors $C$ for different monochromatic SFR estimators (e.g. H$\alpha_{\mathrm{corr}}$) are defined by
\begin{equation}
\mathrm{SFR} ~[\mathrm{M_\sun yr^{-1}}] = C_{\mathrm{H\alpha}} L_{\mathrm{H\alpha, corr}}~ [\rm erg~  s^{-1}].
\label{eq:conversion}
\end{equation}
For \ha\ and FUV values of the parameter $C$ can be derived from stellar population synthesis models by making some suitable simplifying assumption for the star-formation history, the metallicity, and the initial mass function (IMF). This is particularly important for the FUV, which traces longer timescales ($\sim 100$ Myr) than \ha\ ($\sim$~5 Myr, \citealt{Kennicutt2012}). We adopt the $C_{\rm H\alpha}$ value presented by \cite{Calzetti2007}, computed using Starburst99 \cite{Leitherer1999}, a constant star-formation history, age of 100 Myr, solar metallicity and a \cite{Kroupa2001} IMF. Using a \cite{Chabrier2003} IMF changes this factor by less than 5\%, while a \cite{Salpeter1955} IMF requires dividing by a factor of $\sim$ 0.66 \citep{Madau2014}. For reference, this value is within $\approx 2\%$ of the value of $C_{\rm H\alpha}$ suggested by \cite{Murphy2011} and endorsed by \cite{Kennicutt2012}.
For FUV emission, we adopt the value of $C_{\rm FUV}$ recommended by \cite{Leroy2019}, based on the SED fitting results of \cite{Salim2018}, using a \cite{Chabrier2003} IMF. The $C_{\rm FUV}$ factor has a larger dependence on the star formation history, and consequently the position of galaxies in the SFR-$\rm M_\star$ plane (see \citealt{Leroy2019}, their Fig. 25), than $C_{\rm H\alpha}$. The value adopted here is 0.07~dex higher than that recommended by \cite{Kennicutt2012}.


For hybridised SFR estimators (e.g. W4 22~$\rm \mu m$ + H$\rm \alpha_{\mathrm{obs}}$) we define the conversion factors $C$ in a similar fashion as
\begin{equation}
\mathrm{SFR_{H\alpha+ W4}} =  C_{\mathrm{H\alpha}} L_{\mathrm{H\alpha}} + C_{\mathrm{W4}}^{\mathrm{H\alpha}} L_{\mathrm{W4}},
\label{eq:hybrid}
\end{equation}
where, by definition, $C_{\mathrm{H\alpha}}$ is taken to be the same value as the monochromatic estimator (e.q.~\ref{eq:conversion}), so that the sum in Eq. \ref{eq:hybrid} may be interpreted as the combination of an `unobscured' and an `obscured' SFR term. The conversion factor of the W4 $\rm 22~\mu m$ emission is written as $C_{\mathrm{W4}}^{\mathrm{H\alpha}}$ to emphasize that it multiplies the W4 $\rm 22 ~\mu m$ luminosity when hybridised with \ha. In all cases the conversion factors $C$ have units of $\rm M_\sun yr^{-1} /( erg \, s^{-1}) $, which we omit for clarity from now on. We adopt the same approach as \cite{Leroy2019} to construct the UV+IR hybrid SFR calibration, given by
\begin{equation}
\mathrm{SFR_{FUV+ W4}} =  C_{\mathrm{FUV}} L_{\mathrm{FUV}} + C_{\mathrm{W4}}^{\mathrm{FUV}} L_{\mathrm{W4}}.
\label{eq:hybrid2}
\end{equation}

We express $C_{\mathrm{W4}}^{\mathrm{H\alpha}}$ in this form because our main goal is to infer SFRs, but the drawback of this formalism is that it makes the IR-coefficient dependent upon the assumed SFR conversion factor for \ha\ ($C_{\mathrm{H\alpha}}$). In fact, since by equating Equations \ref{eq:conversion} and \ref{eq:hybrid},
$C_{\mathrm{W4}}^{\mathrm{H\alpha}} \propto  C_{\mathrm{H\alpha}}$, our inferred value of $C_{\mathrm{W4}}^{\mathrm{H\alpha}}$ can be directly rescaled to a different assumed $C_{\mathrm{H\alpha}}$. 

We summarise values for coefficients of monochromatic and hybridized SFR estimators from the literature in Table \ref{table:SFR_coeff}. Where relevant, coefficients derived using \textit{Spitzer} MIPS 24 $\rm \mu m$ were revised downward by 0.08 dex to account for the bandpass difference with \textit{WISE} W4 (Sec. \ref{sec:z0mgs}).

\begin{table}[]
	\centering
	\begin{tabular}{llll}
		Coefficient $C$   &  Band(s) & $ \log_{10}(C)$ & reference \\
		$\rm [M_\sun yr^{-1} /(erg \, s^{-1})] $     &                           &                               &           \\  [0.15cm] \hline
		
		$C_{\rm FUV}$  			& FUV		   & $-43.42$  & (1)  \\[0.15cm]
		$C_{\rm W4}^{\rm FUV}$      & FUV+W4       & $-$42.73  & (1) \\[0.15cm]
		$C_{\rm W4}^{\rm FUV}$      & FUV+W4       & $-$42.91  & (2)$^\star$, K12 \\[0.15cm]
		$ C_{\rm H\alpha}$       & \ha\		   &$-$41.26   & (3), K12\\[0.15cm]
		$ C_{\rm W4}^{\rm H\alpha}$  & \ha+W4   	   & $-$42.86  & (3)$^\star$  \\[0.15cm]
		$C_{\rm W4}^{\rm H\alpha}$  & \ha+W4       & $-$43.05  & (4)$^\star$, K12 \\
	\end{tabular}
	\caption{Summary of coefficients from luminosity to SFR, as defined in Eq. \ref{eq:conversion}, \ref{eq:hybrid} and \ref{eq:hybrid2}, taken from the literature and used in this work. $^\star$ implies that the original calibration was derived with \textit{Spitzer} MIPS and has been revised downward by 0.08 dex for application with 22~$\rm \mu m$ WISE W4. \textbf{References} 1: \cite{Leroy2019}, 2: \cite{Hao2011}, 3: \cite{Calzetti2007}, 4: \cite{Kennicutt2009}, K12: \cite{Kennicutt2012}.}
	\label{table:SFR_coeff}
\end{table}

\section{Results}
\label{sec:results}

\subsection{Comparisons among results for existing prescriptions}
\label{sec:compare}

In this section we focus on comparing the SFR obtained from the Balmer decrement-corrected \ha\ against the FUV+W4 prescription of \cite{Leroy2019}. In Fig.~\ref{fig:fig1}, we show SFR maps for three example galaxies from the PHANGS--MUSE sample. NGC~5068 is a low-mass galaxy and the closest object in our sample (15$\arcsec$ = 400~pc). NGC~628 is an archetypal grand-design spiral galaxy, and finally NGC~1566 is a massive barred Seyfert 1 galaxy. 

The comparison between \ha\ and FUV+W4 SFR shows that deviations from equality are spatially correlated with the intensity of star formation activity. In particular, the FUV+W4 SFR is biased high in regions of low \SFR, while the trend is inverted for regions of high \SFR. Such a trend is evident in all three galaxies shown in Fig.~\ref{fig:fig1}, despite clear differences in the \SFR\ distributions, and is also seen in the other PHANGS--MUSE galaxies in our sample. The AGN in NGC~1566 shows excess IR emission, but the region is masked in further analysis as already discussed in Sect. \ref{sec:MUSE}.

\subsubsection{Secondary dependencies}
\label{sec:secondary_dep}

In this section we study how well the SFR estimated from FUV+W4 agrees with the SFR estimated from attenuation-corrected \ha\ as a function of several physical properties (Fig. \ref{fig:fig2}). 

The ratio $\rm \log(SFR_{FUV+W4}/SFR_{H\alpha, corr})$ decreases as a function of both \SFR\ and $\rm sSFR = \Sigma_{SFR}/\Sigma_\star$. On average, our data demonstrate that most regions have a larger $\rm SFR_{FUV+W4}$ than $\rm SFR_{H\alpha}, corr$, but the two estimates are in good agreement at the highest sSFR and \SFR\ values present in our sample, $\rm log(sSFR/yr^{-1}) \gtrsim  -10$ and  $\rm log(\Sigma_{SFR}/M_\odot ~yr^{-1}~ kpc^{-2})>-2$. 

In Panel c, we show the dependence of  $\rm \log(SFR_{FUV+W4}/SFR_{H\alpha}, corr)$ on $E(B-V)$ measured from the Balmer decrement. The vertical scale in this panel has been extended to show in blue the median relation obtained by calculating the SFR using the FUV alone, without any IR correction. Using FUV+IR, we obtain an overestimate of the SFR at low $E(B-V)$, which is driven by an overestimate of the attenuation from the IR term. In fact, at larger values of $E(B-V)$ the IR term provides a correction to the FUV flux  which brings the inferred SFR in agreement with the \ha\ SFR estimate.

No trend is evident between $\rm \log(SFR_{FUV+W4}/SFR_{H\alpha, corr})$ and stellar mass surface density or deprojected galactocentric radius (Fig. \ref{fig:fig2}, panels d \& e) . Considering the light-weighted age of the stellar population (panel f), the $\rm \log(SFR_{FUV+W4}/SFR_{H\alpha, corr})$ ratio increases for older ages, in agreement with the sSFR trend noted above.

We do not find any dependence on distance or on the physical scale corresponding to our 15$\arcsec$ resolution ($\approx 400{-}1400$~pc). In particular, the median trends are consistent within the scatter when considering two sub-samples of galaxies divided according to the median distance of the sample (15.85~Mpc). We have checked that the scatter within individual galaxies is comparable to the sample-wide scatter, indicating small galaxy-to-galaxy variations.

Trends associated with the \ha+W4 calibration using the \cite{Calzetti2007} coefficient are similar and are presented in Fig. \ref{fig:appA1}. The main difference is the substantially smaller scatter, as expected since in this case both SFR estimates rely on \ha\ for the un-attenuated portion.

\begin{figure*}
	\centering
	\includegraphics[width=0.8\textwidth, trim=0 0 0 0, clip]{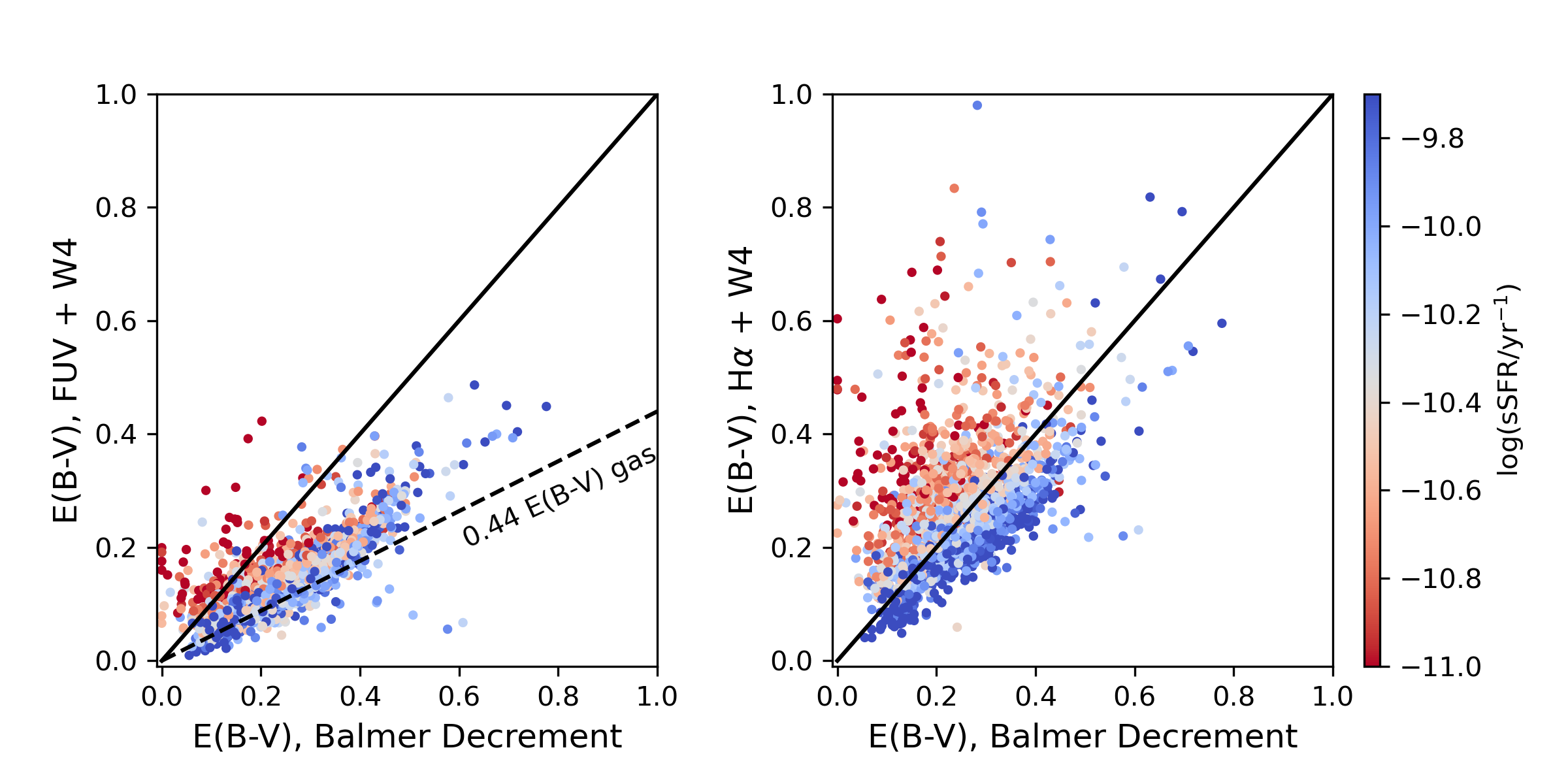}
	\caption{Comparison between $E(B-V)$ estimated via the Balmer Decrement and the same quantity calculated via FUV+W4 (left) and \ha+W4 (right), colour-coded by sSFR. In the left panel we show as a dashed line the ratio between the differential reddening of gas and stars inferred by \cite{Calzetti2000}, $\rm E(B-V)_{stars} = 0.44~ E(B-V)_{gas}$. Solid black lines in both panels represent the one-to-one line. In both panels attenuation is overestimated in regions of low sSFR because of IR cirrus.}
	\label{fig:fig3}
\end{figure*}

\subsubsection{Comparison of the implied E(B-V)}

By equating Equations \ref{eq:hybrid} (or equivalently Equation \ref{eq:hybrid2} for FUV) and \ref{eq:ext_corr_ha} , one can write $E(B-V)$ as a function of the W4/\ha\ or the W4/FUV ratio. In calculating $E(B-V)$ one needs to assume both the hybridisation coefficients (we use Table \ref{table:SFR_coeff}) and an attenuation law to determine $k_\lambda$. In particular, 
\begin{equation}
E(B-V) = \frac{2.5}{k_\lambda} \log{ \left( 1 + \frac{C_{W4}^\lambda L_{W4}} { C_\lambda L_\lambda } \right) },
\label{eq:ebv}
\end{equation}
where $\lambda$ refers to either \ha\ or the FUV band. As discussed in Sect.~\ref{sec:formalism}, we assume a Milky Way-like foreground screen as parameterised by  \cite{O'Donnell1994} for the \ha\ line, and the \cite{Calzetti2001} law for the FUV.

In Fig.~\ref{fig:fig3}, we plot the resulting $E(B-V)$, inferred from FUV + W4 and \ha\ + W4, versus the $E(B-V)_{\mathrm{BD}}$ inferred from the Balmer decrement for the $\sim$ 15$\arcsec$ regions, colour-coded by sSFR. On kiloparsec scales, the attenuation of the stars in the FUV is expected to be lower than that of the ionised gas because nebular emission is more closely linked to the most recent star formation, and therefore to dust. \cite{Calzetti2000} report the differential reddening between the stars and the gas to be a factor of 0.44, which we show as a dashed line in the left panel of Fig.~\ref{fig:fig3} (equivalent to the relation $A_{\rm FUV} = 1.78 ~A_{\rm H\alpha}$). More recent work on nearby galaxies based on MaNGA \citep{Greener2020} supports a somewhat lower ratio, between 0.25 and 0.44. The subset of our regions with the largest sSFR lies close to the latter value, with an average ratio of $E(B-V)_{\mathrm{FUV+W4}}/E(B-V)_{\mathrm{BD}} = 0.52$ for $\rm log(sSFR/  yr^{-1}) > -10 $. On the other hand, regions of low  sSFR, with $\rm log(sSFR/  yr^{-1}) < -10.8 $, show $E(B-V)_{\mathrm{FUV+W4}} > E(B-V)_{\mathrm{BD}} $.



The right panel of Fig.~\ref{fig:fig3} compares $E(B-V)_{\mathrm{H\alpha+W4}}$ with $E(B-V)_{\mathrm{BD}} $ and confirms our findings. For regions with $\rm log(sSFR/ yr^{-1} ) > -10 $, the two estimates of the $E(B-V)$ of the nebular emission agree to within 10\%. At low sSFR, however, $E(B-V)_{\mathrm{H\alpha+W4}}$ is significantly larger, indicating a likely contribution from IR cirrus (see Sect.~\ref{sec:discussion}).

\subsection{A multi-wavelength calibration scheme for SFR at kiloparsec scales}
\label{sec:new_hybrid}

\begin{figure*}
	\centering
	\includegraphics[width=1\textwidth, trim=20 10 20 20, clip]{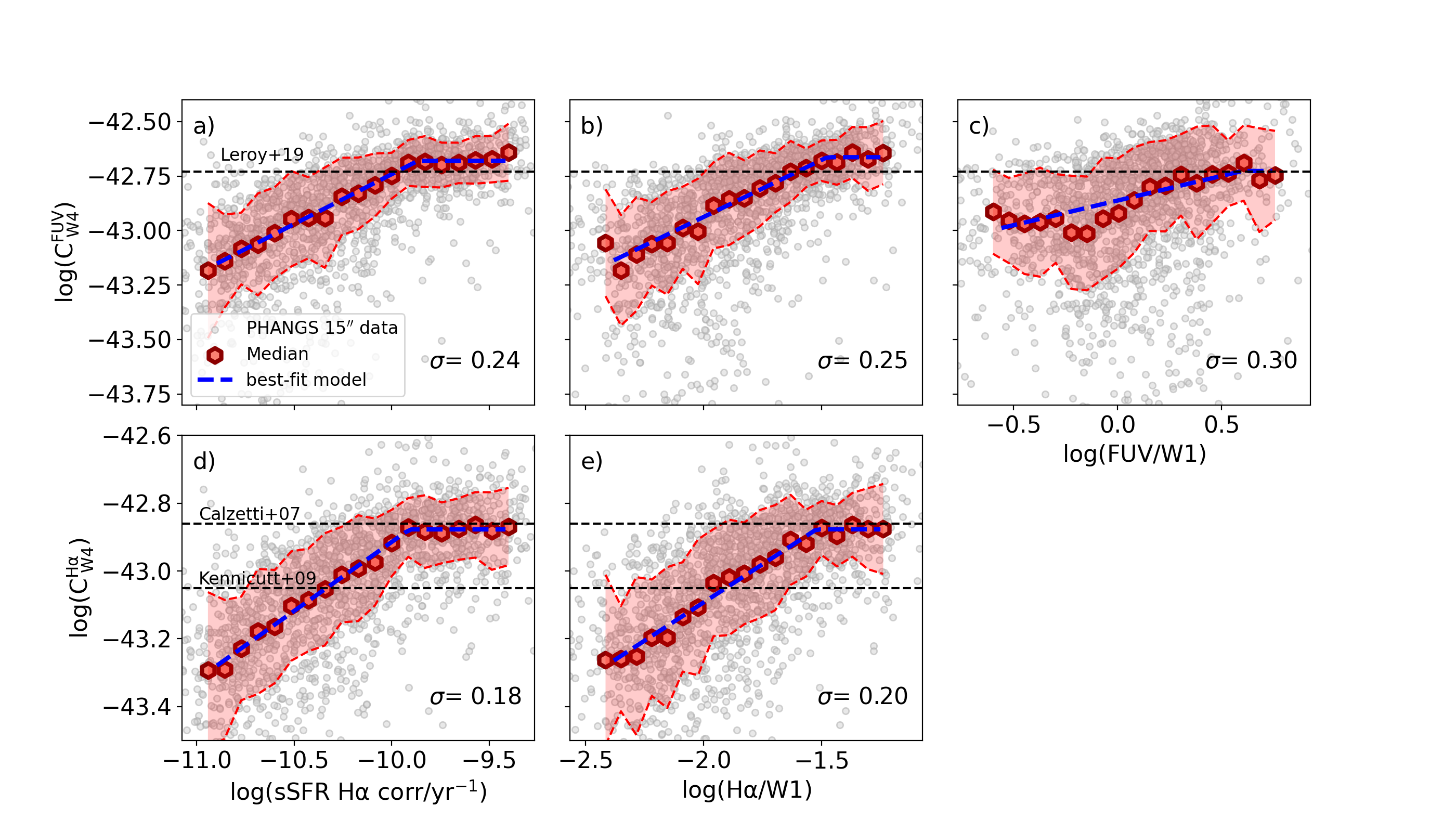}
	\caption{The hybridisation coefficients of W4 with either FUV ($\rm C^{FUV}_{W4}$, Panels a, b, c) or \ha\ ($\rm C^{H\alpha}_{W4}$, Panels d, e) as a function of sSFR (left), $L_{\rm H\alpha}/L_{\rm W1}$ (middle) and $L_{\rm FUV}/L_{\rm W1}$ (right). These coefficients are calibrated to match the SFR obtained from Balmer-decrement-corrected \ha\ for individual 15$\arcsec$ ($\sim$ kpc-scale) regions (grey dots). Red hexagons and shaded areas show the median trends and scatter. Median relations are fitted with a broken power law (equation \ref{eq:hybrid_ratios}) and the best-fit model is shown as a dashed blue line. The scatter $\sigma$ of the data with respect to the best-fit is presented in the bottom-right. Values of $C_{\rm W4}^{\rm FUV}$ and $C_{\rm W4}^{\rm H\alpha}$ from the literature are shown as black dashed horizontal lines. Our data for kpc-scale regions agree well with the literature values at high sSFR, but show a systematic deviation for lower sSFR which is captured well by the power-law model.}
	\label{fig:fig4}
\end{figure*}

In order to bring the FUV+W4 and \ha+W4 SFR into agreement with the star-formation rates obtained from attenuation-corrected \ha, we calibrate the hybridisation coefficients $C_{\rm W4}^{\rm band}$ as a function of physical quantities that trace the ratio of young to old stars \citep[e.g.][]{Boquien2016}. From Sect. \ref{sec:compare} we determined that sSFR, \SFR, and stellar population light-weighted age show well-defined trends, all potentially attributable to IR cirrus. We focus on sSFR, as well as \ha/W1 and FUV/W1 as empirical proxies for sSFR, because they are intensive quantities (unlike \SFR), less model-dependent than the light-weighted age derived from full spectral fitting, and are widely available for nearby galaxies thanks to the combination of \textit{GALEX} and \textit{WISE}.

Assuming the attenuation correction from the Balmer decrement, we use Equation \ref{eq:hybrid} and \ref{eq:conversion} to determine $C_{\rm W4}^{\rm FUV}$ and $C_{\rm W4}^{\rm H\alpha}$, and plot these coefficients as a function of sSFR (Fig. \ref{fig:fig4}, a \& d). Both coefficients show a strong positive correlation with sSFR (Spearman $r$ = 0.71--0.72). We fit the median relation with a broken power law of the form
\begin{equation}
\log{ C_{W4}^{\rm band}} = 
\begin{cases}
 a_0+ a_1 \, \log{Q} & Q<Q_{\rm max},\\
 \log{C_{\rm max}} & Q>Q_{\rm max},\\
\end{cases}   
\label{eq:hybrid_ratios}
\end{equation}
where $Q$ refers to the quantity of interest (e.g. sSFR) and $a_0 =  \log{C_{\rm max}} - a_1 \log{Q_{\rm max}}$ for the function to be continuous at $\log{Q_{\rm max}}$. The best-fit broken power law relations are shown in Fig.~\ref{fig:fig4} as blue dashed lines and the values of the best-fit parameters ($C_{\rm max},  Q_{\rm max}$ and $a_1$) are given in Table \ref{table:hybrid_ratios}. These relations should not be extrapolated past the range populated by our data, we therefore also show in Table \ref{table:hybrid_ratios} the 5$\rm ^{th}$ percentile of the distribution of the quantity of interest, log($Q_{\rm min}$). We do not recommend using the best fits for values lower than log($Q_{\rm min}$).

A broken power law of this form is chosen in order to allow for a constant coefficient at high sSFR. The choice reflects the expectation that the contribution to emission at 22$\rm~\mu m$ from dust heated by old stellar populations becomes negligible in regions dominated by young stars, and therefore the hybridisation coefficients should approach a constant value. In fact, such a flattening is evident in Fig.~\ref{fig:fig4} (Panels a \& d) for both  $C_{\rm W4}^{\rm FUV}$ and $C_{\rm W4}^{\rm H\alpha}$ and $\log{\rm sSFR/yr^{-1}}  \gtrsim -9.9 $. The best-fit value of the coefficients for large sSFR are $C_{\rm W4}^{\rm FUV} = -42.68 \pm 0.05$ and $C_{\rm W4}^{\rm H\alpha} = - 42.88 \pm 0.04$. We consider these values to be our best estimates for the hybridisation coefficients in star-forming regions. The value of $\log{C_{\rm max}}$ for $C_{\rm W4}^{\rm FUV}$ agrees within the error with the estimate by \cite{Leroy2019}, based on galaxy-integrated fluxes ($C_{\rm W4}^{\rm FUV} = -42.73$). Regarding  the $C_{\rm W4}^{\rm H\alpha}$ coefficient, our inferred value of $\log{C_{\rm max}}$ is virtually identical to the value preferred by \cite{Calzetti2007} ($C_{\rm W4}^{\rm H\alpha}= -42.87$, based on star-forming regions) and higher then that of \cite{Kennicutt2009} ($C_{\rm W4}^{\rm H\alpha}= -43.06$, based on galaxy-integrated photometry). 

The parametrisation of $\log{ C_{\rm W4}^{\rm band}}$ in terms of sSFR is instructive, but cannot be used directly to infer the SFR. If a stellar mass measurement is available, however, an iterative approach can be used to determine the SFR. Nonetheless, we also investigate alternative parametrisations in terms of band ratios that can be used as proxies for sSFR. In particular, we consider the luminosity ratios of $\rm H\alpha/W1$ and $\rm FUV/W1$ (Fig. \ref{fig:fig4} middle and right columns). We find that a broken power law of the form of Equation \ref{eq:hybrid_ratios} is a good description for the relation between $C_{\rm W4}^{\rm FUV}$ and $C_{\rm W4}^{\rm H\alpha}$  and the $\rm H\alpha/W1$  luminosity ratio. The scatter between the data and the best-fit models (0.25 dex in the case of $C_{\rm W4}^{\rm FUV}$ and 0.20 dex for $C_{\rm W4}^{\rm H\alpha}$) is comparable to that obtained for the sSFR parametrisation discussed above. Therefore we consider these fits our recommended approach for calculating SFR when \ha\ fluxes are available.

The use of $\rm FUV/W1$ instead of $\rm H\alpha/W1$ leads to larger scatter, and the resulting median relation between the hybridisation coefficients and $\rm FUV/W1$ is less well described by a broken power law. We therefore only recommend the use of $\rm FUV/W1$ when \ha\ data are not available (i.e. only for $C_{\rm W4}^{\rm FUV}$ coefficient). The scatter in the parametrisation in this case goes up to 0.3 dex (Fig.~\ref{fig:fig4}, Panel c).

\begin{table*}[]
	\centering
	\begin{tabular}{lllccccc}
		Band & combined with & as a function of  & $a_1$  & $\log{Q_{\rm max}}$   & $\log{C_{\rm max}}$  & $\log{Q_{\rm min}}$ & $\sigma({\rm log C})$ \\
		\hline
		FUV & W4 &  sSFR 								& $0.46\pm0.14$ & $-9.87\pm0.19$ & $-42.68\pm0.04$  & $-11.00$ & 0.24  \\
		FUV & W4  &  $L_{\rm H\alpha}/L_{\rm W1}$ 	   & $0.52\pm0.15$ & $-1.48\pm0.16$ &  $-42.66\pm0.06$  & $-2.45$ & 0.25  \\
		FUV & W4 &   $L_{\rm FUV}/L_{\rm W1}$  		& $0.23\pm0.14$ &  $0.60\pm0.68$  & $-42.73\pm0.12$  & $-0.64$ & 0.30  \\
		H$\alpha$ & W4   & sSFR & $0.40\pm0.12$  &  $-9.91\pm0.18$ &  $-42.88\pm0.04$ & $-11.00$ & 0.18  \\
		H$\alpha$  & W4  &  $L_{\rm H\alpha}/L_{\rm W1}$     &$ 0.45\pm0.16$  & $ -1.52\pm0.19 $&  $-42.88\pm0.04$ & $-2.45$ & 0.20  \\
	\end{tabular}
	\caption{Coefficients for predicting the hybridization coefficients $C_{\rm W4}^{\rm FUV}$ and $C_{\rm W4}^{\rm H\alpha}$ according to Equation \ref{eq:hybrid_ratios} as a function of quantity $Q$. $a_1$ corresponds to the slope at low $Q$, $\log{C_{\rm max}}$ to the value of the coefficient for $Q>Q_{\rm max}$. $\log{Q_{\rm min}}$ is the 5$\rm ^{th}$ percentile of the distribution of $\log{Q}$ in the PHANGS data. We do not recommend extrapolation below this value. $\sigma({\rm log C})$ refers to the scatter between the data and the best-fit broken power law model.  }
	\label{table:hybrid_ratios}
\end{table*}

Finally, we check the effect of our suggested calibrations on the inferred $E(B-V)$. In particular we focus on the case where W1, W4 and \ha\ data are available, since it is relevant to the galaxies observed by the PHANGS-\ha\ narrow band survey. We calculate $E(B-V)$ via Equation \ref{eq:ebv}, using the calibration of $C_{\rm W4}^{\rm H\alpha}$ as a function of \ha/W1 given by Equation \ref{eq:hybrid_ratios}. Comparing the $E(B-V)$ computed in this fashion with those obtained from the Balmer decrement, we find that they agree very well on average (median offset of  < 0.01 mag) with a scatter of 0.08 mag. This represents a substantial improvement over the situation shown in Fig. \ref{fig:fig3}, where the  $E(B-V)$  obtained using a constant \ha+W4 was on average 0.14 mag larger than the one obtained from the Balmer decrement. An evaluation of the performance of our calibrations is delayed to Sect.~\ref{sec:recommend}.

\subsection{Validity of calibrations as a function of physical scale}
In this section we study the validity of our proposed calibrations as a function of physical scale. In particular, we address the question of whether the \SFR\ estimates obtained at kpc scales are consistent with those measured at 100~pc scales accessible in MUSE surveys of nearby galaxies such as PHANGS-MUSE \citep{Emsellem2022, Pessa2021} or the MUSE Atlas of Discs \citep{Erroz-Ferrer2019}. We then explore whether the calibrations can be applied to spatially-integrated (galaxy-wide) scales, in order to connect to the SFR measurements obtained by large spectroscopic surveys in the nearby Universe.

\subsubsection{Dust extinction from kpc to 100 pc scales}
\label{sec:ha_narrow}

The mixing of regions with different physical conditions within the same resolution element leads to an underestimate of the overall attenuation correction, since the more attenuated regions contribute less to the integrated light in the larger aperture. \cite{ValeAsari2020} showed analytically that, assuming the attenuation law does not vary across a galaxy, a low-resolution measurement of the attenuation-corrected \ha\ luminosity always underestimates the result obtained at high spatial resolution. The magnitude of this underestimation depends in a complex fashion on the spatial distribution of the dust emission and the strength of the correlation between dust attenuation and \ha\ surface brightness. \cite{ValeAsari2020} used MaNGA data to demonstrate that this effect is negligible in practice (3\% on average) when comparing measurements taken on kpc-scale resolution elements to entire galaxies, but they cautioned against extending this result to smaller scales.

\begin{figure}
	\centering
	\includegraphics[width=0.48\textwidth, trim=0 0 0 0, clip]{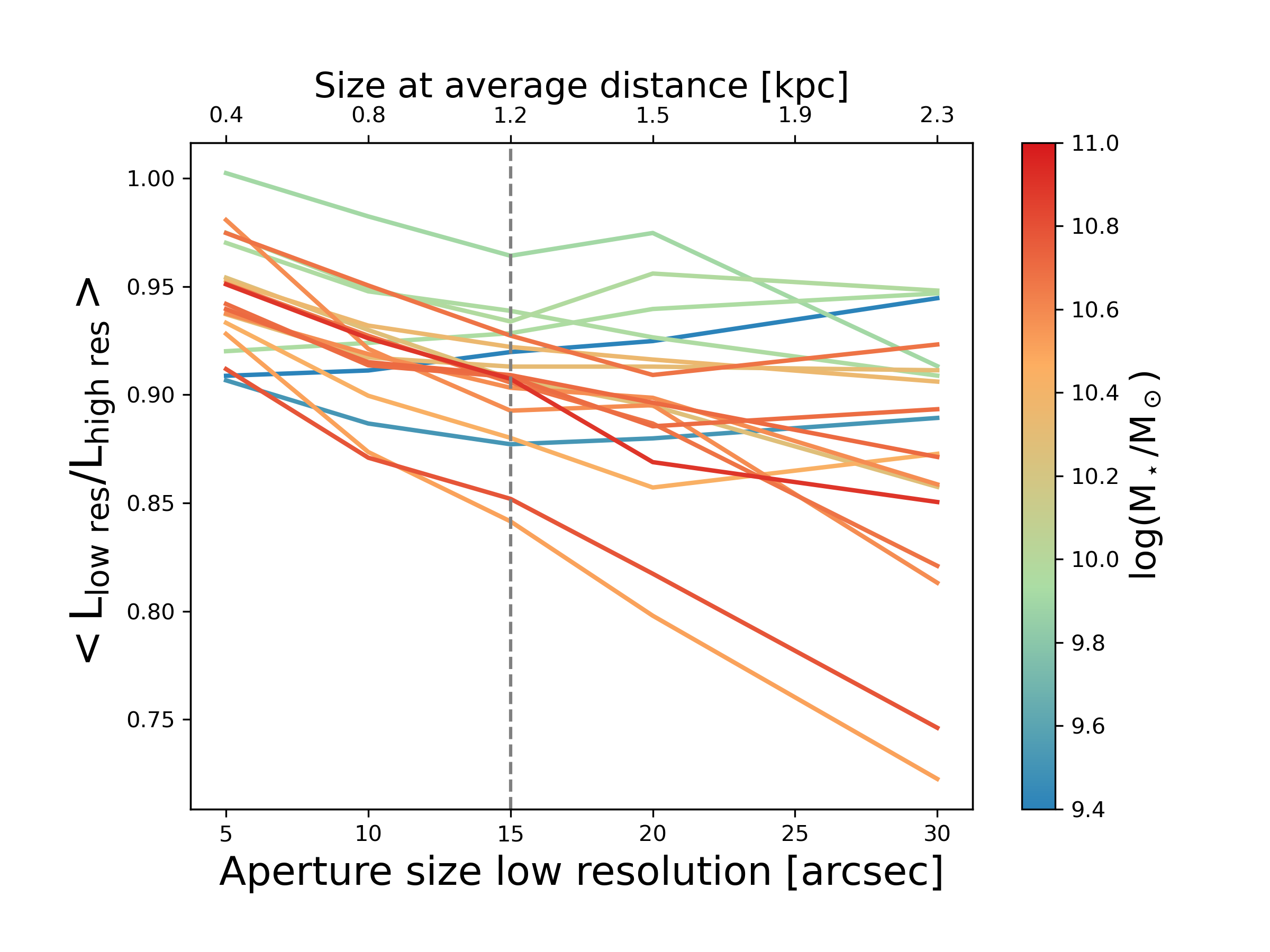}
	\caption{Mean ratio of the attenuation-corrected \ha\ luminosity computed in low-resolution apertures of varying size to the luminosity derived at the highest spatial scales accessible with the PHANGS-MUSE data ($\sim 1\arcsec$), as a function of the low-resolution aperture size. Trends are shown for each of the 19 galaxies in the PHANGS-MUSE sample, colour-coded by total stellar mass. The data for each galaxy is resampled on apertures of 5, 10, 15, 20, and 30$\arcsec$. The alternative x-axis on the top of the plot shows the equivalent physical size in kpc at the median distance of the sample. At 15$\arcsec$ resolution, the resolution of the \textit{WISE} W4 data, the median difference is is 10\% ($\rm <L_{low \ res}/L_{high \ res}> = 0.92$).}
	\label{fig:fig5}
\end{figure}

Here we reconsider the effect of spatial sampling on the attenuation corrections based on the Balmer decrement, focusing on the spatial scales of interest in this work (1--15$\arcsec$, or from $\sim$ 70 to 1100 pc at the average distance of our targets). To do so, we resample the MUSE \ha\ and \hb\ copt maps with pixels of 5$\arcsec$, 10$\arcsec$, 15$\arcsec$ (the sampling adopted in the previous sections to match \textit{WISE} W4), 20$\arcsec$, and 30$\arcsec$. For simplicity we do not convolve the data before resampling, but the effect of the convolution process does not significantly change the outcome of our results. Using these resampled maps we compute the Balmer decrement and the attenuation-corrected \ha\ luminosity ($L_{\rm low \ res}$) in each coarser pixel. We then compare this estimate with that obtained by correcting individual pixels in the original high-resolution maps and summing the attenuation-corrected \ha\ flux ($L_{\rm high \ res}$). We follow this procedure for each galaxy in the sample and compute the average ratio of $L_{\rm low \ res}/L_{\rm high \ res}$ as a function of aperture size. 

Fig. \ref{fig:fig5} demonstrates that the ratio $L_{\rm low \ res}/L_{\rm high \ res}$ varies across the galaxies in our sample, presumably due to differences in the overall dust morphology. At 15$\arcsec$ resolution the underestimation in the attenuation correction with respect to the estimate obtained at the copt (arcsec-resolution) data is on average 10\%, with a 0.1 dex scatter within different regions of individual galaxies. This is a small source of uncertainty in the overall uncertainty budget of the attenuation-corrected \ha.

We checked for a correlation between the slope of the $L_{\rm low \ res}/L_{\rm high \ res}$ versus aperture size and other quantities of galaxies. In particular, we considered the effect of inclination, distance, $\rm M_\star$, SFR and sSFR. The strongest correlation manifests itself with stellar mass (Spearman $r=-0.6$), indicating that low-mass galaxies tend to have flatter trends in this space (as already evident in the colour-coding in Fig.~\ref{fig:fig5}). A mild inverse correlation is also seen with inclination (Spearman $r=-0.32$), where more inclined galaxies have flatter slopes in this space. We have checked the effect of distance on these correlations by rerunning the analysis using apertures of fixed physical size in kpc. We find that the trend with mass persists (Spearman $r=-0.51$), but the trend with inclination is considerably weaker (Spearman $r=-0.09$). 

The result of this exercise is dependent on our approach to the low-S/N area in the arcsec-resolution copt maps. In particular, while \ha\ is detected in nearly 100\% of the pixels of our maps, \hb\ is only detected with S/N $>$ 3 in 85\% of the pixels in each map on average (with some galaxies having up to 50\% of their pixels undetected in \hb\ at the 3 $\sigma$ level). These spaxels contain on average 3\% of the total \ha\ flux, but in some galaxies, this fraction can go up to 10\%. It is important, therefore, to exclude the possibility that the undetected \hb\ hides large amounts of attenuated star formation. 

In order to estimate the importance of dust attenuation in regions where \hb\ is undetected, we make use of the  adaptive binning procedure described in \cite{Belfiore2022} to obtain detection of \hb\ in the low surface brightness regions via binning. This binning procedure was optimised for recovery of emission line fluxes in the diffuse ionised gas, and therefore generates larger bins in regions of faint \ha\ emission (e.g. inter-arm regions). We find that low-surface-brightness pixels lie on the extrapolation of the power law relation between \ha\ and $\rm E(B-V)$ observed at high surface brightness. At \hb\ surface brightness corresponding to a 3 $\sigma$ detection limit in the individual spaxels of our copt maps ($\rm \log{L_{H\beta} / erg \ s^{-1}  kpc^{-2} } = 37.5 $), the median $\rm E(B-V)$ is $\sim$ 0.1~mag ($-$0.1/+0.15). Given this modest level of average attenuation, the relative contribution of \hb\ undetected regions to the attenuation-corrected \ha\ would remain small, and we therefore treat spaxels with S/N in \hb\ less than 3 as having zero attenuation for the purposes of our test. We test the effect of assigning to all the \hb\ undetected spaxel an $\rm E(B-V) = 0.1$~mag and the results do not change substantially.

We conclude that attenuation corrections calculated via \ha\ + W4 (Sec. \ref{sec:new_hybrid}) on kpc scales are in agreement with those derived from arcsec-resolution ($\sim$100 pc physical resolution) Balmer decrement maps. 

\subsubsection{SFR calibration for whole galaxies from integrated UV, IR and \ha\ fluxes}
\label{sec:manga}

We test the validity of our approach for whole galaxies using integrated fluxes for a sample of galaxies from the MaNGA IFS survey. MaNGA is the largest optical IFS survey of the local Universe ($0.01<z<0.15$), offering measurements of the \ha\ and \hb\ fluxes out to at least 1.5 $r$-band effective radii. Aperture corrections in calculating the SFR are negligible, as demonstrated by \cite{Belfiore2018}. In order to test our hybrid SFR estimator on MaNGA data we cross-match the MaNGA catalog from the SDSS Data Release 17 \citep{SDSS_DR17}, and in particular the summary DAPALL catalog file generated by the MaNGA Data Analysis Pipeline \citep{Westfall2019, Belfiore2019a}, with the integrated \textit{GALEX} and \textit{WISE} photometry derived by \cite{Salim2018} as part of GSWLC2\footnote{\url{https://salims.pages.iu.edu/gswlc/}} (\textit{GALEX}-SDSS-\textit{WISE} legacy catalog). We consider the version of the GSWLC2 catalog which contains the deepest available \textit{GALEX} data, denoted as GSWLC2-X2. The MaNGA DR17 catalog contains 10248 galaxies, 86\% of  which have a match in GSLWC-X2. We further restrict the sample to galaxies that are star forming according to the \cite{Kewley2001} criteria in the $\rm [SII]\lambda6717,31/H\alpha$ versus $\rm [OIII]\lambda5007/H\beta$ Baldwin-Phillips-Terlevich \citep{Baldwin1981, Phillips1986} diagnostic diagram, have S/N$> 3$ in \textit{GALEX} FUV, \textit{WISE} W1 and W4 bands and have a semi-axis ratio $b/a >0.4$ (inclination $\sim 66^\circ$), to exclude highly inclined systems.  These cuts lead to a final sample of 1955 galaxies. We also tested the criterion of \cite{Stern2012} for selecting AGN using the \textit{WISE} bands, and find that this leads to the exclusion of only 13 additional galaxies. The results do not depend on whether this additional cut is performed or not.

We first compare the SFR derived by \cite{Salim2018} with those obtained using the attenuation-corrected \ha\ data from MaNGA. \cite{Salim2018} compute SFRs using the CIGALE SED-fitting code applied on UV-optical (\textit{GALEX}-SDSS) data, and constraining the total IR luminosity via the $\rm 22~\mu m$ \textit{WISE} 4 flux (or 12~$\rm \mu m$ \textit{WISE} 3, if the object is not detected in \textit{WISE} 4). Such an approach allows \cite{Salim2018} to also fit for the slope of the attenuation curve without having access to data in the far-IR. The SFR estimates from \cite{Salim2018} agree well with those we obtain from attenuation-corrected \ha, with a relative scatter of 0.16 dex and an offset of 0.09 dex. The CIGALE-derived SFR estimates are larger than those obtained from attenuation-corrected 
H$\alpha$ and the offset is independent of the sSFR. The reason for the discrepancy is unclear,
but may include differences in the treatment of the attenuation law.

\begin{figure*}
	\centering
	\includegraphics[width=1\textwidth, trim=20 10 20 20, clip]{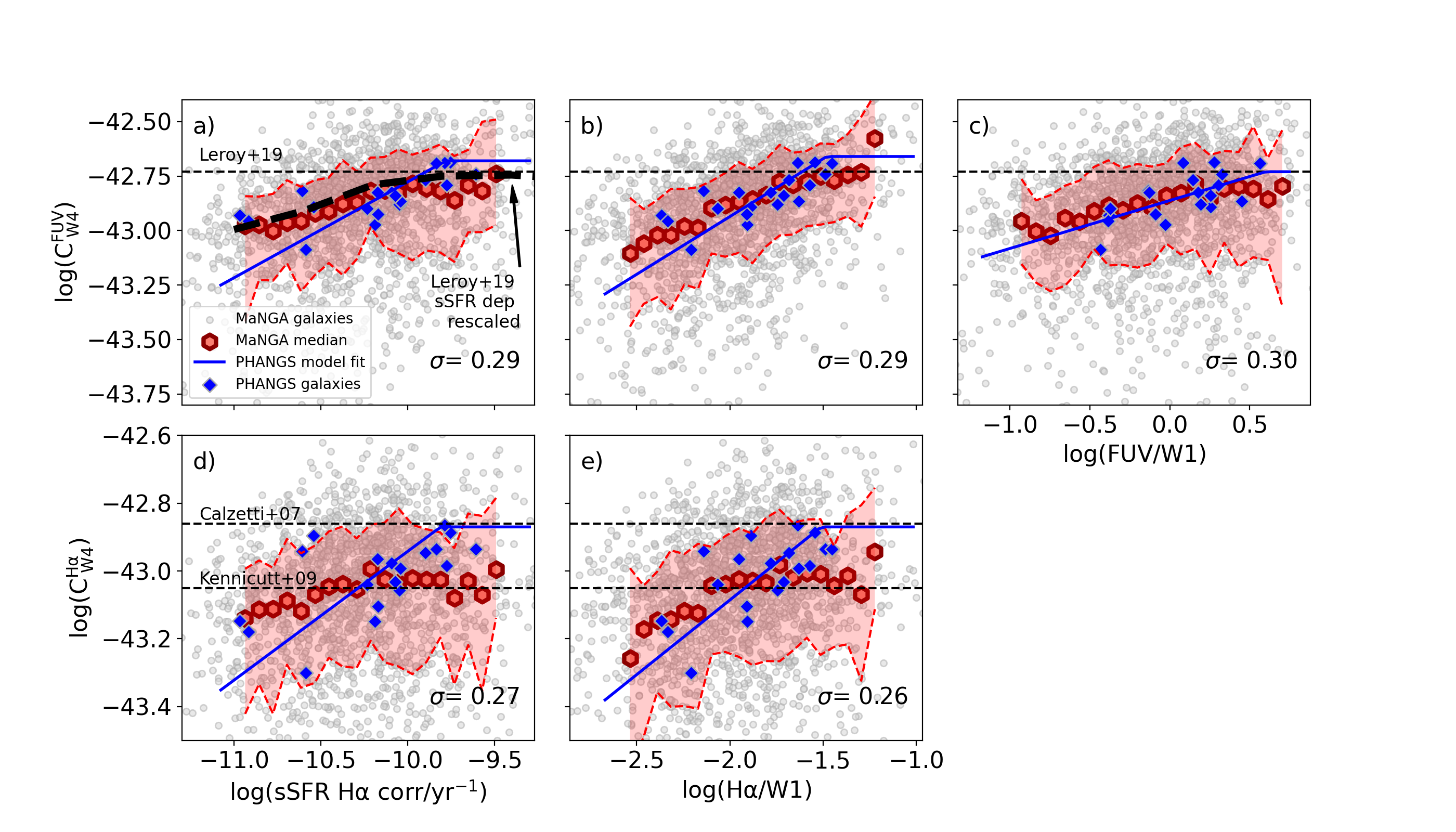}
	\caption{The hybridization coefficients of W4 with either FUV (Panels a, b, c) or \ha\ (Panels d, e) as a function of sSFR (left), $L_{\rm H\alpha}/L_{\rm W1}$ (middle) and $L_{\rm FUV}/L_{\rm W1}$ (right) for the MaNGA sample (using galaxy-integrated fluxes). Best-fit relations obtained for the PHANGS 15$\arcsec$ data (Fig.~\ref{fig:fig4}) are shown in blue. Blue diamonds show the results obtained considering the integrated fluxes from the 19 PHANGS galaxies. Values of $C_{\rm W4}^{\rm FUV}$ and $C_{\rm W4}^{\rm H\alpha}$ from the literature are shown as black dashed horizontal lines and labelled in the figure (see text for additional detail). The scatter $\sigma$ of the data with respect to the best-fit is presented in the bottom-right. The MaNGA data show similar trends as the resolved regions in PHANGS galaxies, even though in general it displays flatter slopes and does not reach the same values of $C$ at the highest sSFR. }
	\label{fig:fig6}
\end{figure*}

To test the applicability of our derived hybrid calibrations, we repeat the analysis in Sect. \ref{sec:new_hybrid} using the MaNGA/GSWLC data. In Fig.~\ref{fig:fig6} we show the dependencies of $C_{\rm W4}^{\rm FUV}$ and $C_{\rm W4}^{\rm H\alpha}$ on sSFR, $\rm L_{H\alpha}/L_{W1}$ and $\rm L_{FUV}/L_{W1}$ for the MaNGA data (grey points). All the physical quantities involved are computed in the same way as for the PHANGS/z0MGS data to allow for a direct comparison. We find that the trends present in the MaNGA data go in the same direction as those observed in PHANGS: galaxies with lower sSFRs show lower values of the $C$ coefficients. At high sSFR, however, the MaNGA data tend to lie below the best-fit model from PHANGS (shown in blue in Fig.~\ref{fig:fig6}). This discrepancy is more evident for the \ha+IR calibration than for the FUV+IR one. 

The results from MaNGA are consistent with the value given by \cite{Kennicutt2009}, who also used integrated measurements of galaxies for their calibration, but lower than the best estimate from \cite{Calzetti2007}, who used data on smaller spatial scales. Depending on a proxy for the sSFR, \cite{Boquien2016} find an amplitude of the variation of $\rm C_{FUV}^{MIPS 24}$ of a factor 2--3 on kpc-sized regions in nearby galaxies, a value that is very similar to what is found for integrated MaNGA galaxies here. \cite{Leroy2019} find a similar sSFR-dependent trend for $C_{\rm W4}^{\rm FUV}$ using the SFR from \cite{Salim2018}. In Fig.~\ref{fig:fig6} we show the median sSFR-dependent trend obtained by \cite{Leroy2019} (from their Table 5), scaled by the 0.09~dex offset discussed above to bring the \cite{Salim2018} and \ha\ attenuation-corrected SFR in agreement. After performing this correction we are in excellent agreement with \cite{Leroy2019}.

The trends in Fig.~\ref{fig:fig6} show that the contribution of old stellar populations to the heating of the dust is significant even at high sSFR because of the unavoidable mixing of emission from a variety of regions (e.g. arm and interarm) when considering galaxies as a whole. This mixing implies a lower $C$ coefficient than that measured for kpc-scale regions of high sSFR. The slope of the trend at lower sSFR is also shallower for MaNGA than for PHANGS, which can also be qualitatively explained by an averaging effect. Finally, we show as blue diamonds the results obtained on integrated scales for the 19 galaxies in the PHANGS-MUSE sample. In particular, for each galaxy we integrate the relevant fluxes within the MUSE mosaic coverage, taking the masks into account. We find that the PHANGS sample agrees within the scatter with the general population of local galaxies from MaNGA.

The MaNGA sample leads to relations that show a larger scatter than for PHANGS regions, highlighting the wider range of conditions found in the overall galaxy population. The trends do not change if we make a more aggressive inclination cut, e.g. $b/a >0.6$ ($i < 53^\circ$), although a stricter cut decreases the number of systems with low FUV/W1 ratios. In fact, we find that the larger sampling of low FUV/W1 values in MaNGA with respect to PHANGS is primarily due to inclination effects. Because of the larger scatter, we do not provide new fits to the median relations based on the MaNGA data. We evaluate the effect on the estimated SFR of using the PHANGS-based corrections on the MaNGA data in the next section.

\begin{table*}[]
	\centering
	\begin{tabular}{l c c c c c c c c}
		band	 & combined with & $C_{W4}$ dependent on & recommended & reference & \multicolumn{2}{c}{PHANGS} & \multicolumn{2}{c}{MaNGA} \\
		\hline
		& &  & & & median  & scatter & median  &scatter  \\
		& & & &  & offset [dex] & [dex] & offset [dex] & [dex] \\
		\hline
		FUV & W4 & constant& no &(1)  & 0.08 &  0.17 & ~0.12 & 0.25 \\
		FUV & W4 & $\log(\rm FUV/W1)$ & yes & this work &0.003 &  0.16 & $-$0.007 & 0.22 \\
		FUV & W4 & $\log(\rm H\alpha/W1)$ &no& this work& 0.003 &  0.13 & $-$0.02 & 0.20 \\
		H$\alpha$ & W4  & constant & no & (2) or (3) &0.07 &  0.12 & ~0.003 & 0.18 \\
		H$\alpha$ & W4  & $\log(\rm H\alpha/W1)$ & yes &this work & $-$0.003 &  0.08 & $-$0.002 & 0.14 \\
	\end{tabular}
	\caption{Quality assessment of the SFR recipes discussed in this work in terms of their median offset and scatter with respect to \ha\ attenuation-corrected SFR. Literature recipes with constant coefficients are also shown for comparison. In particular, for the case of \ha+W4, we use the value from (2) for PHANGS and the value from (3) for MaNGA. Notes: (1) \cite{Leroy2019}, (2) \cite{Calzetti2007}, (3) \cite{Kennicutt2009} }
	\label{table:sfr_scatter}
\end{table*}

\section{Discussion}
\label{sec:discussion}
\subsection{Recommendations on estimating SFR from hybrid indicators}
\label{sec:recommend}

\begin{figure*}
	\centering
	\includegraphics[width=1\textwidth, trim=20 10 20 20, clip]{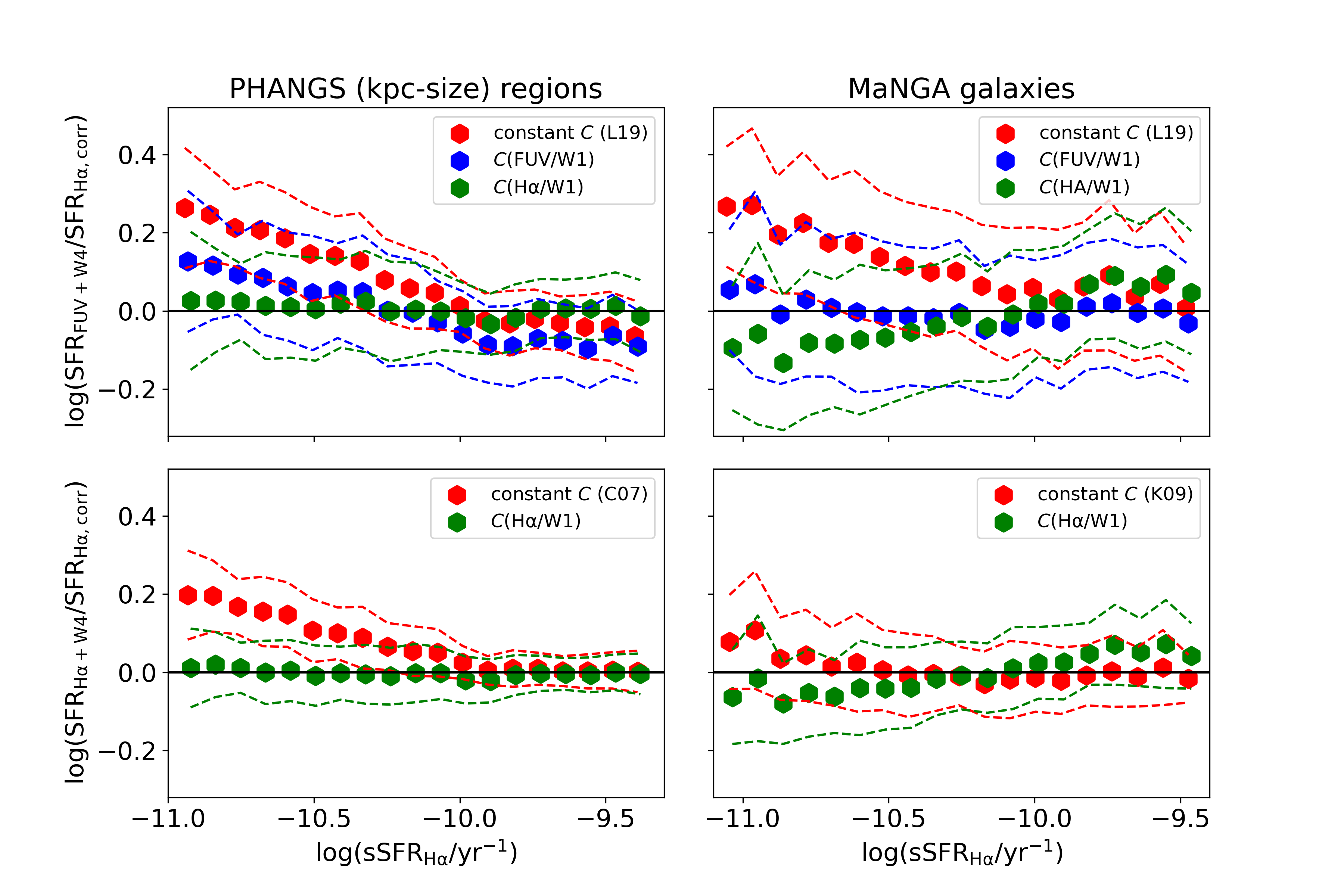}
	\caption{Comparison of the SFR estimated from hybrid tracers (FUV+W4 and \ha+W4) and from attenuation-corrected \ha\ as a function of sSFR. We consider different calibrations, including the use of constant $C$ coefficients (red points, \citealt{Leroy2019}, L19 for the FUV+W4 and \citealt{Calzetti2007}, C07, and \citealt{Kennicutt2009}, K09, for \ha+W4, see Table \ref{table:SFR_coeff}), and PHANGS-based calibrations with $C$(FUV/W1) (blue points) and $C$(\ha/W1) (green points) derived in this work. Left panels show the results for PHANGS (kpc-scale) 15$\arcsec$ regions, while the right panel show the result for MaNGA galaxies. Dashed lines show the 1~$\sigma$ scatter while hexagons the median relations. The mean offset and scatter for each calibration, for both PHANGS and MaNGA, is presented in Table \ref{table:sfr_scatter}.}
	\label{fig:fig7}
\end{figure*}

We evaluate the success of our kpc-scale calibrations based on PHANGS data (Eq. \ref{eq:hybrid_ratios}) by comparing the median offset and scatter between the hybrid estimators and the SFR inferred from attenuation-corrected \ha. In Fig.~\ref{fig:fig7} we show the median log ratio and the scatter of the hybrid SFR estimators considered in Sect.~\ref{sec:new_hybrid} as a function of sSFR for both kpc-scale regions in PHANGS and for galaxy-integrated fluxes from MaNGA. We also consider the usage of constant $C$ coefficients taken from the literature. For the case of the \ha+W4 calibration, we use the \cite{Calzetti2007} value for PHANGS and the the \cite{Kennicutt2009} value for MaNGA, in order to match each coefficient to the range of spatial scales it was computed for. In fact, our results confirm both coefficients are accurate and their difference is due to the different level of cirrus contamination affecting typical galaxies as a function of spatial scale. The median value of the offset and scatter for each data set and calibration strategy are provided in Table \ref{table:sfr_scatter}.

For both calibrations and both data sets, the use of constant coefficients (red points in Fig.~\ref{fig:fig7}) leads to an overestimation of the SFR using the hybrid calibrations at low sSFR levels. We therefore recommend against the use of constant coefficients, especially when investigators are interested in comparing regions or galaxies of different sSFRs. We find, however, that the coefficient are roughly constant for $\rm log(sSFR/yr^{-1}) > -9.9$.

The \ha\ + W4 calibration using a $C$ coefficient scaling with \ha/W1 (green points in bottom panel, Fig.~\ref{fig:fig7}) leads to the least scatter in both PHANGS and MaNGA (0.08 and 0.14~dex, respectively). This calibration performs extremely well for PHANGS regions because of its negligible residual dependence on sSFR. In case of the integrated galaxies from MaNGA, it shows a slight sSFR-dependent tilt, underestimating the SFR at low sSFR and overestimating at high sSFR. This tilt is the result of a flatter slope in the $C_{\rm W4}^{H\alpha}$ versus \ha/W1 plane shown by the MaNGA data with respect to PHANGS shown in Panel e of Fig.~\ref{fig:fig6}. 

When \ha\ data are not available, we recommend to use the FUV+W4 calibration with a $C$ coefficient depending on FUV/W1 (blue points in Fig.~\ref{fig:fig7}). This calibration also successfully removes the mean offset between the hybrid SFR and the \ha-based SFR in both MaNGA and PHANGS. Despite being calibrated on the PHANGS data, it performs better on MaNGA, where it more successfully removes the residual sSFR dependence. 

In summary, our best-effort SFR indicators require the use of \textit{WISE} W1 band in addition to the canonical bands used for hybridization: FUV (or \ha) and W4. 
When compared to Balmer-decrement corrected \ha\ SFR, our calibrations using \textit{WISE} W4 lead to a scatter smaller than 0.2~dex. Since \textit{WISE} is an all-sky survey, our prescriptions can be readily applied to a large sample of local galaxies (e.g. z0MGS, \citealt{Leroy2019}).

The use of closely related bands, including IRAC1, J, H, K band instead of W1 and MIPS24 or \textit{JWST} MIRI F2100W (21 $\rm \mu m$) band instead of W4 is expected to lead to comparable results if appropriate conversion factors are applied. In anticipation of a comparison with upcoming results from \textit{JWST} we note the mean ratio $\log{(L_{\rm F2100W}/L_{\rm W4})} = -0.10$ obtained by \cite{LEROY1_PHANGSJWST} for four galaxies in the PHANGS-MUSE sample (NGC~0628, NGC~1365, NGC~7496, IC~5332). The use of the WISE W3 band (centered at $\rm 11.6~ \mu m$) would lead to additional systematic uncertainty in the SFR estimation because this band includes strong features from polycyclic aromatic hydrocarbon (PAH). The flux ratio between PAH-tracing bands and 24 $\rm \mu m$ emission is not constant, and may depend systematically on metallicity and the intensity of the radiation field \citep{Calzetti2007, Lee2013}. We nonetheless calculate the ratio between W3 and W4 fluxes in our data to allow a first-order correction and obtain a mean ratio of $\log{(L_{\rm W3}/L_{\rm W4})} = 0.08$ for our PHANGS kpc-scale regions. As already shown in \cite{Leroy2019}, however, the $C_{\rm W3}^{\rm FUV}$ factor shows significant changes across the $M_\star$-SFR plane (their Fig. 21), therefore pointing to the importance of additional physics in addition to the cirrus contamination we aim to correct for in this work.

\subsection{Caveats and future prospects}
\label{sec:prospects}

Our calibrations derived from the small PHANGS galaxy sample are found to generalise well to the larger population of main sequence galaxies probed on integrated scales by the MaNGA survey. However, our sample does not contain dwarf galaxies or ultra-luminous infrared galaxies. PHANGS-MUSE data is also limited to the inner star-forming disc and has limited coverage of the \textsc{Hi}-dominated outer parts ($r_{\rm gal} \gtrsim 0.5~r_{25}$) of disk galaxies. In particular, PHANGS-MUSE  does not cover extended UV discs \citep{Thilker2007} outside the optical radius at all. \cite{Tomicic2019}, for example, study regions in the \textsc{Hi}-dominated disc of M31 with IFS and find conversion factors for the IR term (both $\rm C_{\rm W4}^{\rm H\alpha}$ and $C_{\rm W4}^{\rm FUV}$) a factor of 5-8 times larger than the ones we obtain here. They attribute this difference to the high inclination of M31 ($\sim 77^\circ$, \citealt{Walterbos1988}) and the flaring of the \textsc{Hi} disc. Within the limits of our data, we do not find any hint of such a radial trend, but it is reasonable to expect an increased mid-IR cirrus contribution in the presence of a large diffuse ISM component. We highlight future extension of this analysis to the \textsc{Hi}-dominated outer parts of galaxies as an important future direction.


The impact of the old stellar populations on the calibration of hybrid SFR calibrators on resolved scale has previously been discussed by \cite{Boquien2016}. They studied a sample of eight nearby galaxies and used SED fitting based on UV-to-FIR imaging to derive the attenuation in the FUV, and therefore attenuation-corrected SFR. In agreement with our work, they find a trend between $\rm C_{W4}^{FUV}$ and sSFR and other band ratios used as sSFR proxies (including FUV - 3.6 $\mu m$). In particular, they find a flattening in the relation between $\rm C_{W4}^{FUV}$ and sSFR at $\rm sSFR \sim -9.75$, comparable to what is found here. However, their conversion factors are systematically higher than the ones derived here by $\sim $0.3 dex, perhaps due to differing assumptions on the attenuation curve as they allowed for curves steeper than a bona fide starburst curve. This shows the dependence of hybrid estimators on indirect but necessary assumptions on the attenuation curve as we discuss below.

The calibrations presented in this paper are consistent with the large body of literature presenting SFR calibrations based, directly or indirectly, on attenuation-corrected recombination line fluxes \citep{Calzetti2007, Kennicutt2009, Hao2011, Catalan-Torrecilla2015}. They represent an update, extending the validity of the calibrations to smaller physical scales with respect to the recipes presented in the review of \cite{Kennicutt2012}. Our calibrations are also in agreement with the SFRs measured based on radio continuum by \cite{Murphy2012} for a sample of nuclear and extra-nuclear star-forming regions.

The calibrations presented in this work are subject to the systematic uncertainties associated with attenuation corrections via the Balmer decrement and with using \ha\ as a tracer of star formation. Regarding the dust correction, uncertainties are related to the temperature (and density) dependence of the intrinsic \ha/\hb\ ratio and to the use of a Milky-Way-like extinction law to model attenuation of the nebular lines. The intrinsic Balmer decrement changes only weakly with temperature, increasing from 2.76 for 5000~K to 3.05 at 20000~K, therefore causing changes of less than 0.02~mag in the derived $E(B-V)$. The attenuation law (and in particular the assumed value of $R_V$) for the nebular component represents a more significant systematic uncertainty. Assuming a \cite{Calzetti2000} attenuation law for the nebular continuum does not substantially change our conclusions. However, radiative transfer models predict a change in the slope of the attenuation law with dust optical depth \citep{Chevallard2013, Salim2020, Tacchella2022} and dust attenuation curves have been shown to vary within and between galaxies at all redshifts \citep[e.g.][and many others]{Salmon2016, Buat2018, Salim2018, Decleir2019, Reddy2020, Boquien2022}. For example, implementing the attenuation and slope of the curve presented by \cite{Tacchella2022} implies substantially higher values of $A_{\rm H\alpha}$ than using a Milky Way law, especially for the more attenuated regions, and would therefore lead a substantial recalibration of the IR coefficients presented here. More detailed observational work is required in order to confirm changes in the slope of the nebular attenuation law as a function of optical depth. 

The conversion between attenuation-corrected \ha\ flux and SFR is affected by additional systematic uncertainties, including the escape of ionising radiation from galaxies, the absorption of ionising photons from dust, the stochastic sampling of the IMF in low-mass clusters \citep{DaSilva2014}, and to some extent the variability of star formation at low SFR. The fraction of Lyman continuum photons absorbed by dust is found to be significant in nearby galaxies and clusters, between 30-50\% \citep{Inoue2001, Iglesias-Paramo2004}, in agreement with some theoretical models of radiatative transfer \citep{Kado-Fong2020, Tacchella2022}. However, substantial uncertainty affect these estimates, and in detailed models of dusty \hii\ regions dust competes with hydrogen only for high values of the ionisation parameter (the ratio of ionising photons to hydrogen atoms), a regime only relevant to young, compact \hii regions. Potential dust absorption of ionising photons is not taken into account by our calibration, and would imply a change in the $C_{\rm H\alpha}$ factor. For example, assuming the fraction of Lyman continuum photons directly absorbed by dust to be 30\%,  $C_{\rm H\alpha}$ would need to be increased by 0.23 dex. Stochastic sampling of the IMF in low-mass clusters will, on the other hand, cause a bias (and  increased scatter) at the lowest SFR levels, but should only have a limited effect on our kpc-scale data of the inner regions of discs.

At kpc resolution it is not possible to distinguish individual \hii\ regions from the diffuse emission surrounding them. The nature of the diffuse emission at \ha\ and 22~$\rm \mu m$, however, represents one of the main systematic uncertainties to the determination of SFR on even smaller spatial scales. The diffuse \ha\ emission (also known as diffuse ionised gas) is directly powered by ionising radiation leaking from \hii\ regions \citep{Ferguson1996, Zurita2000, Haffner2009, Belfiore2022}. In particular, \cite{Belfiore2022} studied the same sample of galaxies used in this work and showed that the DIG is consistent with being powered by leaking radiation, with a mean free path of $\sim 2$~kpc, therefore having negligible impact on the SFR estimates on kpc scales.

Diffuse emission at 22~$\rm \mu m$ results from the convolution of the dust distribution with the local radiation field. Therefore it is potentially sensitive to both heating due to the radiation field from old stars \citep{Leroy2008, Verley2009, Kennicutt2009, Leroy2012} and the presence of an extended dust-bearing gas reservoir, especially atomic gas, mixed with the old stellar population \citep[e.g.,][]{Leroy2012}. 

The calibrations we present in this work effectively model this spatially unresolved background emission by making the IR calibration coefficient dependent on an sSFR-like quantity. Higher-resolution IR maps, capable of resolving dust heated by \hii\ regions from the diffuse component, are needed to directly test this interpretation. While \textit{Spitzer} 24~$\rm \mu m$ offers cloud-scale $\sim$ 100 pc resolution in the Local Group and in galaxies $\lesssim 4$ Mpc (e.g. \citealt{Helou2004, Verley2009,Faesi2014}), these targets probe only a limited set of environments. The PHANGS-JWST program will address this shortcoming by providing sub-arcsec-resolution 21~$\rm \mu m$ mapping of the 19 PHANGS-MUSE galaxies using the MIRI instrument on board \textit{JWST}. In combination with arcsec-resolution mapping of \ha\ and \hb\ from MUSE, and FUV maps at $\sim 2\arcsec$ resolution from \textit{AstroSat} (Hassani et al. in preparation), this combined data set will shed new light on the nature of diffuse IR emission, small-scale behaviour of dust attenuation in star-forming discs, and the statistics of fully-embedded regions \citep{Prescott2007}.

As discussed of above, a complementary future direction will be to extend this analysis to the outer, atomic gas-dominated parts of galaxies. In these regions, we might expect much of the dust (i.e., the dust mixed with diffuse atomic gas) to reside in an extended distribution rather than being concentrated towards the peaks of star formation. Such a situation could produce increased infrared cirrus emission even when the average sSFR remains high and so will represent an important test of the generality of our prescriptions.

\section{Summary and conclusions}
\label{sec:summary}

In this work we use \ha\ maps corrected for attenuation using the Balmer decrement to calibrate multi-wavelength hybrid SFR recipes for use in nearby galaxies at kpc-scale resolution. We make use of a sample of 19 galaxies observed with optical IFS as part of the PHANGS-MUSE program, and combine this data with FUV imaging from \textit{GALEX} and mid-IR imaging at 22 $\rm \mu m$ from \textit{WISE} W4. We focus on calibrating two commonly used hybrid recipes: FUV+W4 and \ha+W4. We further assess the reliability of the resulting calibrations as a function of physical scale. We summarise our main results below.
\begin{itemize}
    \item Calibrations of FUV+W4 and \ha+W4 from the literature overestimate the SFR obtained from attenuation-corrected \ha\ in regions of low sSFR, or equivalently low \SFR, old light-weighed stellar age or low E(B$-$V). We attribute this trend to the presence of `IR cirrus': diffuse IR emission originating from dust heated by the interstellar radiation field and not associated with star formation.
    \item In order to correct for IR cirrus we propose to modify the calibration term in front of the IR term in the hybrid SFR recipes and make it a function of band ratios which correlate with sSFR. We present broken power law fits (equation \ref{eq:hybrid_ratios}) for the dependence of $\rm C_{FUV}^{W4}$ on \ha/W1 and FUV/W1 and of $\rm C_{H\alpha}^{W4}$ on \ha/W1 in Table \ref{table:hybrid_ratios}. These calibrations lead to SFR which agree with those obtained from attenuation-corrected \ha\ with negligible bias and scatter of less than 0.16 dex.
    \item The SFR obtained on kpc scales with these recipes agrees well with the estimates from attenuation-corrected \ha\ calculated on scales of 100~pc using the MUSE IFS data at its native resolution. In particular, we find that based on our sample of 19 star-forming galaxies the lower-resolution SFR is underestimated by less 10\% on average. 
    \item We test the validity of our calibration on the scales of entire galaxies by using attenuation-corrected \ha\ data for star-forming galaxies from the MaNGA IFS survey. We observe comparable trends in terms of the sSFR dependence of the IR coefficient, albeit with quantitative differences compatible with the inevitable mixing of regions with various levels of star formation activity when considering galaxies as whole. Nonetheless, we find that the calibrations derived from PHANGS kpc-scale data perform well with MaNGA, leading to a scatter smaller than 0.2~dex when comparing hybrid and \ha\ attenuation-corrected SFRs.
    \item At high sSFR our calibrations are in good agreement with the coefficients recommended by the \cite{Kennicutt2012} review. In particular, for the \ha+W4 calibration the kpc-scale analysis of the PHANGS-MUSE data is comparable with the results of \cite{Calzetti2007} at high sSFR, and the MaNGA analysis of flux-integrated galaxies is compatible with the coefficient derived by \cite{Kennicutt2009}. However, our calibrations will lead to more accurate SFR measurements than obtained using constant coefficients from the literature when considering samples of regions or galaxies sampling a wide range of sSFR.
\end{itemize}

\begin{acknowledgements}
	
	This work has been carried out as part of the PHANGS collaboration.
	Based on observations collected at the European Southern Observatory under ESO programmes 094.C-0623 (PI: Kreckel), 095.C-0473,  098.C-0484 (PI: Blanc), 1100.B-0651 (PHANGS-MUSE; PI: Schinnerer), as well as 094.B-0321 (MAGNUM; PI: Marconi), 099.B-0242, 0100.B-0116, 098.B-0551 (MAD; PI: Carollo) and 097.B-0640 (TIMER; PI: Gadotti).
	Science-level MUSE mosaicked datacubes and high-level analysis products are provided via the ESO archive phase 3 interface\footnote{\url{https://archive.eso.org/scienceportal/home?data_collection=PHANGS}}. A full description of the the first PHANGS data release is presented in \cite{Emsellem2022}.
	The work of AKL was partially supported by the National Science Foundation (NSF) under Grants No.~1615105, 1615109, and 1653300.
	The work of JS is partially supported by the Natural Sciences and Engineering Research Council of Canada (NSERC) through a Canadian Institute for Theoretical Astrophysics (CITA) National Fellowship. ATB would like to acknowledge funding from the European Research Council (ERC) under the European Union’s Horizon 2020 research and innovation programme (grant agreement No.726384/Empire, PI Bigiel).
	MB gratefully acknowledges support by the ANID BASAL project FB210003 and from the FONDECYT regular grant 1211000. EC acknowledge support from ANID Basal projects ACE210002 and FB210003. OE and KK gratefully acknowledge funding from the German Research Foundation (DFG) in the form of an Emmy Noether Research Group (grant No. KR4598/2-1, PI Kreckel).
	CE gratefully acknowledges funding from the Deutsche Forschungsgemeinschaft (DFG) Sachbeihilfe, grant number BI1546/3-1.
	SCOG and RSK thank for funding from the Heidelberg Cluster of Excellence EXC 2181 (Project-ID 390900948) `STRUCTURES', supported by the German Excellence Strategy, from the ERC in the Synergy Drant `ECOGAL' (project ID 855130),  from DFG via the Collaborative Research Center (SFB 881, Project-ID 138713538) `The Milky Way System' (subprojects A1, B1, B2, B8), and from the German Ministry for Economic Affairs and Climate Action for funding in  project `MAINN' (funding ID 50OO2206).
    KG is supported by the Australian Research Council through the Discovery Early Career Researcher Award (DECRA) Fellowship DE220100766 funded by the Australian Government. MQ acknowledges support from the Spanish grant PID2019-106027GA-C44, funded by MCIN/AEI/10.13039/501100011033.  
    PSB acknowledges financial support from the Spanish Ministry of Science, Innovation and Universities under grant number PID2019-107427GB-C31. ES and TGW acknowledge funding from the European Research Council (ERC) under the European Union’s Horizon 2020 research and innovation programme (grant agreement No. 694343, PI Schinnerer).

\end{acknowledgements}

%
\bibliographystyle{aa} 
\bibliography{library7, personal} 


\begin{appendix}





\end{appendix}

\end{document}